\documentclass[twocolumn,secnumarabic,amssymb, nobibnotes, aps, prd]{revtex4-1}

\usepackage{graphicx}
\usepackage{hyperref}
\hypersetup{colorlinks,citecolor=blue}
\usepackage{amsmath}
\usepackage{xcolor}

\begin{document}
\color{red}
\title{Conformally symmetric traversable wormholes in modified teleparallel gravity}%

\author{Ksh. Newton Singh}%
\email[Email:]{ntnphy@gmail.com}
\affiliation{Department of Physics, National Defence Academy, Khadakwasla, Pune-411023, India. \\
Department of Mathematics, Jadavpur University, Kolkata-700032, India}

\author{Ayan Banerjee }
\email[Email:]{ayan\_7575@yahoo.co.in}
\affiliation{Department of Mathematics, Jadavpur University, Kolkata-700032, India}

\author{Farook Rahaman}%
\email[Email:]{rahaman@associates.iucaa.in}
\affiliation{Department of Mathematics, Jadavpur University, Kolkata-700032, India}

\author{M. K. Jasim}
\email[Email:]{mahmoodkhalid@unizwa.edu.om}
\affiliation {Department of Mathematical and Physical Sciences, University of Nizwa, Nizwa, Sultanate of Oman.}

\date{today}%

\begin{abstract}
In this paper, we consider wormhole geometries in the context of  teleparallel equivalent of general relativity (TEGR) as well as $f(T)$ gravity.  The TEGR is an alternative geometrical formulation of Einstein's general relativity, where  modified teleparallel gravity or $f(T)$ gravity has been invoked as an alternative approach for explaining an accelerated expansion of the universe. We present the analytical solutions under the assumption of spherical symmetry and the existence of a conformal Killing vectors to proceed a more systematic approach in searching for exact wormhole solutions. More preciously, the existence of a conformal symmetry places restrictions on the model. Considering the field equations with a diagonal tetrad and anisotropic distribution of the fluid, we study the properties of traversable wormholes in TEGR that violates the weak and the null energy conditions at the throat and its vicinity. In the second part, wormhole solutions are constructed in the framework of $f(T)$ gravity, where $T$ represents torsion scalar. As a consistency check, we also discuss the behavior of energy conditions with a viable power-law $f(T)$ model and the corresponding shape functions. In addition, a wide variety of solutions are deduced
by considering a linear equation of state relating the density and pressure, for the isotropic and anisotropic pressure, independently of the shape functions, and various phantom wormhole geometries are explored.

\end{abstract}


\maketitle

\section{Introduction}\label{sec1}

Wormholes are popular tools in science fiction which act as a tunnel-like structure connecting different universes or widely separated regions lie in the same universe. These geometrical model can be considered as a way for rapid interstellar travel,  time machines and warp drives. In this direction investigation led by Einstein and Rosen \cite{Einstein} in the middle of the 1935s, where they constructed an elementary particle model represented by a \textit{bridge} linking two identical sheets: Einstein-Rosen bridge (ERB).  This mathematical representation of physical space was an unsuccessful particle model. Twenty years later Wheeler \cite{Wheeler} interested in topological issues in GR, which he denoted ``gravitational-electromagnetic entity" in short geons. These were considered as a 
 configuration of the gravitational field, possibly coupled to other zero-mass fields such as massless neutrinos \cite{Wheeler1} or the electromagnetic field \cite{Hartle}. Dubbing Einstein-Rosen bridges, Wheeler \cite{Wheeler2} sought a way that particles  would emerge from a kind of spacetime foam connecting different regions of spacetime at the Planck scale. However, the terminology of ``wormhole'' was first coined by Wheeler \cite{whel} in 1957. In the late 1962s, Fuller and Wheeler \cite{Fuller1962} were able to prove that {\bf ERB} would collapse instantly upon formation. In fact, ERB is a non-traversable wormhole, even by a photon.

Modern interest in wormholes are mainly based on the seminal work by Morris and Thorne \cite{Morris:1988cz} and subsequently Morris, Thorne and Yurtsever \cite{Morris:1988tu}. This issue was investigated after introducing a static spherically symmetric metric with the desired structures and then recovered the matter fields through Einstein's equation. The character of wormhole is asymptotically flat with a constant or variable radius which  depends on its configuration. These authors have showed that wormholes can be traversable provided that they are supported by ``exotic" matter with a minimal surface area linked to satisfy flare-out condition, which is called throat of the wormhole. It turns out that the energy-momentum tensor to violate the null energy conditions (NEC) called exotic matter; in fact wormhole solutions violate all of the standard energy conditions \cite{Visser}.  However, this type of matter sounds to be unusual in general relativity (GR), but from a quantum gravitational perspective, they are seen as a natural consequence if the topology of spacetime fluctuates in time \cite{Wheeler}. As a consequence, it is important and useful to minimize the violation of the energy conditions or reduce the encounter of exotic matter at the throat.  The energy conditions of dynamical wormholes in general relativity was explored in 
Ref. \cite{Ori1993,Ori1994,Wang:1995nj}.

A notable result was the `volume integral quantifier' proposed by Visser \textit{et al.} \cite{Visser:2003yf}. They showed that the amount of exotic matter can be made infinitesimally small by choosing the geometry of the wormhole in a very specific and appropriate way. In the same direction another interesting proposal came from Kuhfittig \cite{Kuhfittig:2002ur,Kuhfittig:1999ur} by imposing a condition on $b'(r)$ to be close to one at the throat. It has been demonstrated that evolving wormhole without violation of the weak energy condition (WEC) could exist within classical GR \cite{Kar:1994tz,Kar:1995ss}. They consider a metric with a conformal time-dependent factor,  whose spacelike sections are $R \times S^2$ with a wormhole metric. The thin-shell formalism is another approach to minimize the exotic matter, where the exotic matter is  concentrated at the throat \cite{Visser:1989kh,Visser:1989kg}. 

It goes without saying that violation of NEC is an unavoidable consequence within GR. However, in the context of modified gravity and higher dimensional theories, it was shown that the normal matter threading the wormhole satisfies all of the energy conditions. In this regard, the study of wormhole solutions in modified theories of gravity enchanted researchers in avoiding the presence of these non standard fluids. More precisely, all known modifications to Einstein gravity introduce new degrees of freedom in the gravitational sector. In particular, it was shown that wormhole throats can be constructed without the presence of exotic matter in $f(R)$ gravity \cite{Lobo:2009ip}. In this context, wormhole geometries have been studied assuming different fluids with specific shape functions and examined the validity of energy conditions in \cite{Mazharimousavi:2012xv,Pavlovic:2014gba,Sharif:2018jdj,DeBenedictis:2012qz}. Author of Ref. \cite{MontelongoGarcia:2010xd} studied the wormhole geometries supported by a non-minimal curvature-matter coupling.  It turns out that the matter threading the solution satisfies the null energy condition. This type of solutions were also found in Einstein-Gauss-Bonnet theory  \cite{Bhawal:1992sz,Mehdizadeh:2015jra,Kanti:2011yv,Maeda:2008nz},  Born-Infeld gravity \cite{Shaikh:2018yku} and  Lovelock gravity \cite{Dehghani:2009zza,Zangeneh:2015jda,Matulich:2011ct}.
In the curvature-matter coupled theory, $f(R,T)$ gravity, exact solutions were found 
\cite{Moraes:2017mir,Elizalde:2018frj,Elizalde:2018arz,Banerjee:2019wjj}.

Motivated by the above discussion, our aim is to find wormhole solutions in $f(T)$ gravity. Inspired by the formulation of higher-order gravity theories, such as $f(R)$ theories, `Teleparallel Equivalent of General Relativity' (TEGR) has been generalized to $f(T)$ gravity. Interestingly, TEGR \cite{Unzicker:2005tz,Shirafuji:1995xc,Maluf:2013gaa,Okolow:2013ifa} is a gravity theory based on spacetime torsion, replacing a zero torsion Levi-Civita connection by a zero curvature Weitzenb$\ddot{\text{o}}$ck connection with the vierbein as a fundamental tool. The Weitzenb$\ddot{\text{o}}$ck linear
connection furnishes a null Riemann curvature tensor, and  characterize a globally flat space-time endowed with a non-zero torsion tensor. It is important to 
mention that this connection is metric-compatible, and the dynamics of the theory is based only on the torsion. This is one of the main differences between GR and TEGR, and comparing with GR, TEGR has some advantages \cite{Ortin} over the conventional formulation.

In analogy to the above theory, the so-called $f(T)$ gravity has been introduced as a straight forward modification of Teleparallel gravity by changing in the TEGR action. These $f(T)$ gravity models, where $T$ is the torsion scalar, the Lagrangian is taken to be a non-linear function of the TEGR Lagrangian $T$ \cite{Ferraro:2006jd,Ferraro:2008ey,Cai:2015emx}. This construction crucially depended on an appropriate ansatz for the tetrad field. In contrast to this, the equations of motion of the torsion-based $f(T )$ theory involve only the usual second order derivatives of the tetrad fields. However, the curvature-based $f(R)$ gravity leads to fourth order derivative of the metric in the resulting equations of motion. Subsequently, in the pure tetrad formalism (absence of the spin connection), $f(T)$ gravity exhibits violation of the local Lorentz invariance \cite{Cai:2015emx} (for more review see \cite{Ferraro:2006jd}). Other  potentially helpful approaches in this respect must be explored, as the proposals of \emph{covariant} formulation of teleparallel gravities where the spin connection is taken different from zero 
\cite{Krssak:2015oua,Golovnev:2017dox,Bejarano:2019fii}. Thus, the choice of the tetrad always a sensitive issue in $f(T)$ theory and different tetrads might give rise to different solutions. The good and bad tetrads in $f(T)$ gravity has been widely studied in \cite{Tamanini:2012hg}. In this context, $f(T)$ theories can potentially be used to explain the late-time cosmic accelerating expansion without invoking dark energy \cite{Wu:2010xk,Wu:2010mn,Myrzakulov:2010vz,Karami:2013rda}, cosmological perturbations \cite{Chen:2010va,Dent:2011zz}, spherically symmetric solutions \cite{Wang:2011xf}, solar system constraints \cite{Iorio:2012cm}, and so on. 

Additionally, the application of $f(T)$ gravity are not restricted only for cosmological solutions, but there are wide range of application in astrophysics also. In this theory, static and spherically symmetric  solutions were considered \cite{Deliduman:2011ga,Wu:2011xa,Nashed:uja}.  By analyzing different tetrads in detail, Boehmer \emph{et al} \cite{Boehmer:2011gw} proved the existence of relativistic stars in $f(T)$ gravity and explicitly constructed several classes of static perfect fluid solutions. Furthermore, in a recent paper  \cite{Singh:2019ykp}, we  proposed a new approach to find Einstein's cluster solution that mimicking the behaviors of compact star.  In fact compact stars have been  theoretically modelled within the frame work of $f(T)$ gravity (for reviews see Ref. 
\cite{Das:2015gwa,Chanda:2019hyh,Abbas:2015xia}).

The main aim of this paper is to present a class of wormhole solutions with a diagonal tetrad and assuming different hypotheses for their matter content in both TEGR and $f(T)$ gravity. In \cite{Bohmer:2011si} an off-diagonal tetrad has considered to explore traversable wormhole geometries are supported by $f(T)$ gravity. It was demonstrated that obtained solution satisfies the  weak and the null energy conditions at the throat and its vicinity. In this line of direction several solutions have been thoroughly analyzed for wormholes \cite{Jamil:2012ti,Lin:2019tyw,Sharif:2013exa}. It should be stressed that, dynamical wormhole in $f(T)$ gravity was found in \cite{Sharif:2013lya}.

Here, we focus on a new class of traversable wormholes where the spacetime is assumed to be spherically symmetric and to possess a conformal symmetry. Nevertheless, an exact solutions of traversable wormholes were found under the assumption of \textit{non-static} conformal symmetry \cite{Boehmer:2007md}. To be more precise, conformal symmetry gives a natural link between geometry and matter through the Einstein field equations. Our work reveals the  feature of conformal Killing operator $\mathcal{L}$ associate with the metric $g$  is a linear mapping from the space $\mathcal{J}(\bf \xi)$ of vector fields on ${\bf \xi}$, which yield
 \begin{equation}\label{eq0}
\mathcal{L}_\xi  g_{ik}=\psi  g_{ik}, ~~~\text{where}~~~  \xi \in \mathcal{J}(\bf \xi),
\end{equation}
where $ \psi$ is the conformal factor and the metric $g$ is conformally mapped onto itself along $\xi$. Note that for $ \psi$ is zero,  we refer to this as a \emph{true} Killing vector and the metric is completely invariant as it is dragged along the curves with that true Killing vector as their tangent vector. In favor of this mathematical technique was applied by Herrera \textit {et al} \cite{Herrera1984,Herrera1985} and in \cite{Maartens} to show that for a one-parameter group of conformal motions, the EoS is uniquely determined by the Einstein equations. Further, strange quark stars with respect to one class of admissible transformations was explored in Ref \cite{Mak2004}.

Meanwhile, this approach has been utilized very successfully in wormhole geometry \cite{Kuhfittig:2015cea,Rahaman:2013ywa,Bhar:2016vdn,Sharif:2016aom}. The main motivation comes  from a recent article by us \cite{Banerjee:2019wjj}, where  wormholes were found under the assumption of spherically symmetric and to possess a conformal symmetry in $f(R,T)$ gravity. This paper is outlined in the following manner:  After the introduction in Section \ref{sec1}, we briefly review the basics of the $f(T)$ gravity model in Section \ref{sec2}. In Section \ref{sec3}, we give an overview about the general geometries and constraints of traversable wormholes. Section \ref{sec4}, is devoted to study the field equations of the $f(T)$ theory with a linear and power-law model of $f(T)$ functions, respectively. In Section \ref{sec5}, exact general solutions are deduced using static conformal symmetries for both TEGR and $f(T)$ gravity. In Section \ref{sec6}, we explore the wormhole geometries in TEGR by assuming suitable conditions.  In the case of TEGR wormholes  matter violates the null and weak energy conditions at the throat and its vicinity. While, in Section \ref{sec7}, we devoted to explore
the wormhole solutions by assuming a power-law $f(T )$ model as
well as different shape functions. Summary and conclusions are reported in Section \ref{sec8}.

\section{Teleparallel gravity and its modifications: Basic equations and action}\label{sec2}

Before starting our considerations on $f(T)$ gravity and its astrophysical realization, it is useful to briefly review the $f(T)$ gravitational paradigm. The notation is as follows: Greek indices $\mu$, $\nu$,...run over the coordinate space-time and lower case Latin indices $i$, $j$,... run over the tangent space-time. We begin by recalling that the dynamical variables in teleparallel gravity are the vierbein or tetrad fields, $e^{i}_{\mu}$, which satisfy
\begin{eqnarray}\label{a1}
e^i_\mu e^\nu_i=\delta^\mu_\nu ~~ \text{and} ~~ e^i_\mu e^\mu_j=\delta^j_i,
\end{eqnarray} 
where $\delta$ is the Kronecker tensor.  Thus, the spacetime
metric tensor and the tetrads are related by
\begin{eqnarray}\label{a2}
g_{\mu \nu} (x)= \eta_{ij} e^i_\mu(x)e^j_\nu(x),
\end{eqnarray}
where $\eta_{ij}$ is the Minkowski metric of the tangent space with the form of $\eta_{ij}= \text{diag}~(1,-1,-1,-1)$. The metric $g$ is used to raise and lower
coordinate indices and $\eta$ raises and lowers frame indices.

Since, Teleparallel gravity carries a fundamental distinction from curvature based descriptions of gravity. Instead of using the torsionless Levi-Civita connection in GR, one uses the Weitzenb$\ddot{\text{o}}$ck connection, which is given by
\begin{eqnarray}\label{a3}
\widetilde{T}^\sigma_{\mu \nu}= e_i^\sigma \partial_{\nu}e^i_\mu= -e^i_\mu \partial_{\nu}e_i^\sigma.
\end{eqnarray}
With the above consideration, the covariant derivative, $D_\mu$, of the tetrad fields 
\begin{eqnarray}\label{a4}
D_\mu e_{\nu}^i \equiv \partial_{\mu}e^i_\nu- \widetilde{T}^\sigma_{\mu \nu}e^i_\sigma,
\end{eqnarray}
vanishes identically, leads to a vanishing scalar curvature but non-zero torsion.

Now, introducing the torsion and contorsion tensors, to clarify the interrelations between Weitzenb$\ddot{\text{o}}$ck and Levi-Civita connections, which are 
\begin{eqnarray}\label{a5}
T^\sigma_{\mu \nu} &=& \widetilde{T}^\sigma_{ \nu \mu}-\widetilde{T}^\sigma_{\mu \nu}
=e^\sigma_i \left(\partial_{\nu}e^i_\mu-\partial_{\mu}e^i_\nu   \right)  ,\\
K^{\mu \nu}_\sigma &\equiv & T^\sigma_{\mu \nu}-\widetilde{T}^\sigma_{\mu \nu} 
= {1 \over 2} \Big(T^{\mu \nu}_\sigma + T^{\nu \mu}_\sigma - T^{\mu \nu}_\sigma \Big),\label{a6}
\end{eqnarray}
respectively. Furthermore, the super-potential tensor relates the torsion and contorsion tensors, as follows 
\begin{eqnarray}\label{a7}
S^{\mu \nu}_\sigma &=& K^{\mu \nu}_\sigma-\delta^\nu_\sigma ~T^{\alpha \mu}_\alpha+\delta^\mu_\sigma~ T^{\alpha \nu}_\alpha,
\end{eqnarray}
Finally, we define the torsion scalar $T$, as
\begin{eqnarray}\label{a8}
T \equiv T^\sigma_{\mu \nu}S^{\mu \nu}_\sigma,
\end{eqnarray}
which used in the action and varied in terms of the vierbeins give rise to the same equations with GR. Thus, the torsion-based variant of the theory is known as the teleparallel equivalent of General Relativity (TEGR). Analogous to $f(R)$ gravity, Teleparallel gravity has been extended by constructing gravitational Lagrangians to a function $f(T )$  of a torsion scalar $T$.

Therefore, the corresponding action of $f(T )$  gravity reads as (with geometrized
units $c = G = 1$)
\begin{eqnarray}\label{a9}
S= {1\over 16  \pi} \int e f(T)~ d^4x+ \int e \mathcal{L} ~d^4x,
\end{eqnarray}
where $e$ is the determinant of $e^{i}_{\mu}$ and $\mathcal{L}$ is the matter Lagrangian.
 
Now, varying the resultant action with respect to the tetrads $e^{i}_{\mu}$, one obtains the following field equation for $f(T)$ gravity:
\begin{eqnarray}\label{a10}
 S^{\mu \nu}_i f_{TT}~\partial_{\mu}T + e^{-1} \partial_{\mu}(eS^{\mu \nu}_i)~f_{T} \nonumber \\
 - T^\sigma_{\mu i} S^{\nu \mu}_\sigma f_{T}- 
 {1\over 4} e^\nu_i f = -4\pi \mathcal{T}^\nu_i .
\end{eqnarray}
where $f_{T} =df(T)/dT$ and $f_{TT}=d^2f(T)/dT^2$, and the tensor $\mathcal{T}^\nu_i$ represents the energy-momentum tensor of the matter source $\mathcal{L}$.  When $f(T ) = T$, the action is the same as in TEGR, and $f(T ) = T-2\Lambda$, the equations of motion (\ref{a10}) are the same as that of the Teleparallel theory with a cosmological constant, and this is dynamically equivalent to the GR. 

Since, the field equation (\ref{a10}) appears very different from Einstein's equations due to partial derivatives and tetrad components.

\section{Traversability conditions and general remarks for wormholes} \label{sec3}
The spacetime ansatz for seeking traversable wormholes are described by a static and spherically symmetric metric which is in the usual spherical $(t, r, \theta, \phi)$ coordinates, and the corresponding line element can be written as \cite{Morris:1988cz},
\begin{equation}\label{a11}
ds^{2}=e^{\nu(r)}dt^{2}-\left(1-{b(r) \over r}\right)^{-1} dr^{2}-r^{2}\left(d\theta^{2}+\sin^{2}\theta d\phi^{2} \right),
\end{equation}
where $\nu(r)$ and $b(r)$ are the redshift and the shape functions, respectively.  The function $b(r)$ in Eq. (\ref{a11}) is called the shape function, since it represents the spatial shape of the wormhole. The shape function $b(r)$ should obey the boundary condition $b(r = r_0$) = $r_0$ at the throat $r_0$ where $r_0 \leq r \leq \infty$. In order to describe the wormhole solution, the shape function must satisfy the \textit{flaring-out} condition that can be obtained from the embedding calculation, and reads
\begin{equation}\label{a12}
\frac{b(r)-rb^{\prime}(r)}{b^2(r)}>0.
\end{equation}
Mathematically the above condition can be also written in a short way, namely,
$b^{\prime}(r_0) < 1$ at the throat $r = r_0$. Since, the geometry is static and spherically symmetric, we assume that $\nu(r)$ should be finite everywhere in order to avoid the presence of an event horizon \cite{Morris:1988cz}. The condition $1 - b(r)/r \geq 0$ is also imposed. Another important criterion is  the proper radial distance $\ell(r)$, defined as
\begin{equation}\label{a13}
\ell (r) = \pm \int^{r}_{r_0}{\frac{dr}{\sqrt{1-\frac{b(r)}{r}}}},
\end{equation}
is required to be finite everywhere. Thus, the proper distance decreases from the upper universe $\ell = +\infty$ to the throat of the wormhole $\ell$ and then from $\ell = 0$ to $\ell = -\infty$ in the lower universe. Moreover, `$\ell$' should  greater than or equal to the coordinate  distance, i.e. $ \mid \ell (r) \mid$ $\geq$ $r-r_0$; the $\pm$ signs denote the upper and lower parts of the wormhole which are connected by the wormhole throat.  The embedding surface of the wormhole can be observed by determining the embedding surface $z(r)$ at a fixed time $t=\mbox{const}$ and $\theta = \pi/2$. With this constraint the metric of Eq. (\ref{a11}) becomes,
\begin{equation}
z'(r) = \pm {1 \over \sqrt{r/b(r)-1}}. \label{dem}
\end{equation}
Considering the conformal symmetry, the above equation turns out to be
\begin{equation}
z(r) = \pm \int {c_3 \over \psi} \left(1-{\psi^2 \over c_3^2} \right)~ dr.  \label{emb}
\end{equation}

In the present situation, we consider the matter is described by an anisotropic stress-energy tensor of the form
\begin{eqnarray}\label{a14}
\mathcal{T}^\nu_i &=& (\rho+p_t)u^\nu u_i + p_t g^\nu_i + (p_r-p_t)\chi_i \chi^\nu, \label{eq8}
\end{eqnarray}
where $u_\nu$ is the four-velocity and $\chi_\nu$ is the unit spacelike vector in the radial direction. In the following expression $\rho(r)$ is the energy density, $p_r = p_r(r)$ and $p_t = p_r(r)$ are the radial and transverse pressures, respectively. If matter is considered to be isotropic then $p_r=p_t$. Throughout the discussion  prime denotes the derivative with respect to the radial coordinate $r$.

\section{Wormhole solutions in different forms of $f(T)$ } \label{sec4}. Since, by considering different forms of $f(T)$'s we  arrive at different field equations with the choice of a set of diagonal tetrads. Here, we will consider two classes of solution, (i) linear function in $f(T)$ i.e. TEGR, and (ii) a viable power-law form of the $f(T)$ model.

\subsection{Field equations in teleparallel gravity $f(T)=aT+B$}

In order to compute the field equations, we employ the following diagonal tetrad \cite{Abbas:2015zua,Momeni:2016oai},
\begin{equation}\label{eq10}
[e^i_\mu]= \text{diag}(e^{\nu/2},~(1-b/r)^{-1/2},~r,~r \sin \theta),
 \end{equation}
 and its determinant is $|e^i_\mu|=r^2 \sin \theta ~ e^{\nu/2}(1-b/r)^{-1/2}$. 
 As a consequence, the torsion scalar becomes
 \begin{eqnarray}
T(r) &=& {2 \over r} \left(1-{b(r) \over r}\right) \left(\nu'(r)+{1 \over r} \right), \label{eq11}
\end{eqnarray}
where the prime denotes the derivative with respect to $r$. 

Since we know that the off diagonal components of the field equations vanish in the static case of GR, whereas for $f(T)$ gravity there exists a $(r, \theta)$ component which gives an extra equation $T' f_{,TT}=0$. But this equation does not appear in the corresponding curvature-based equations of motion. It rises from the specific choice of tetrad in spite of that they are diagonal. According to the Ref. \cite{Tamanini:2012hg}, this equation leads to  satisfy either $f_{,TT} = 0$ or $T' = 0$, where the former reduces the theory to TEGR. In what follows, the choice of $f_{,TT} = 0$ leads to the following linear model $f(T)=aT+B$ \cite{Abbas:2015zua}, which are of physical interest in this context. 
 
 Now, inserting the vierbein choice (\ref{eq10}) into the field equations (\ref{a10}) we obtain the set of equations for an anisotropic fluid as
\begin{eqnarray}
4\pi \rho &=& \frac{1}{4} \left(\frac{2 a b'}{r^2}+B\right), \label{eq18}\\
4\pi p_r &=& \frac{r^2 \left(2 a \nu '-B r\right)-2 b \left(a r \nu '+a\right)}{4 r^3} ,  \label{eq19} \\
4\pi p_t &=& \frac{1}{8 r^3} \Big[r \big(a b' (a \nu '+2)+r \big\{2 a r \nu ''+a \nu ' (r \nu '+2) \nonumber \\
&& \hspace{-0.8 cm}  -2 B r\big\}\big)-a b \left\{\nu ' (a+r^2 \nu '+2 r)+2 r^2 \nu ''+2\right\} \Big], \label{eq20}
\end{eqnarray}
where $\rho$ is the energy density with  $p_r$ and $p_t$ are the radial and tangential pressure of the matter sector, respectively.

\subsection{Field equations in $f(T)=aT^2+B$}
Here, we will study for the choice of a set of diagonal tetrads with a particular power-law form of $f(T)$ model i.e. $f(T)=aT^2+B$, where $a$ and $B$ are constants. It has been shown that the power-law inflation model can easily accommodate with the regular thermal expanding history including the radiation and cold dark matter dominated phases. Utilizing the model along with Eq. (\ref{a11}), we obtain the following expression
\begin{eqnarray}
4\pi \rho &=& \frac{1}{4 r^6} \Big[ r^2 \left(24 a b'-36 a+B r^4\right)-24 a r b \left(b'-3\right) \nonumber \\
&& -36 a b^2 \Big], \label{eq21}\\
4\pi p_r &=& \frac{27 a b^2}{r^6}-\frac{48 a b}{r^5}+\frac{21 a}{r^4}-\frac{B}{4}, \label{eq22}\\ 
4\pi p_t &=& \frac{1}{4 r^6} \Big[r^2 \left(60 a-24 a b'-B r^4\right)+24 a r b \nonumber \\
&& \left(b'-4\right)+36 a b^2 \Big].\label{eq23}
\end{eqnarray}
Since, the choice of $f(T)$ in this case do not satisfy $f_{,TT}=0$ and $T'=0$. So, the field equations are not corresponding to TEGR.

Notice that above field equations for both models give three independent equations with five unknown quantities i.e. $\rho(r)$, $p_r(r)$, $p_t(r)$, $\nu(r)$ and $b(r)$. Therefore, system of equations is under-determined, and we shall reduce the number of unknown functions by assuming suitable conditions.

\section{Conformal killing vectors}\label{sec5}

Despite the success of numerical computation, exact solutions are still important in GR as well as modified gravity, because they allow a \textit{global} acceptance without specifying the choice of parameters and initial conditions. In this sprite  conformal symmetries provide important insight and information into the general properties of self-gravitating matter configurations. Guided by the above motivations we assume that the  static and spherically symmetric spacetime admit a Conformal Motion. According to Ref. \cite{Herrera1984,Maartens},  we simplify the problem and build up its basic mathematical structure. 

In general, Conformal Motion (CM) is a map $M \rightarrow M$ such that the metric $g$ of the spacetime transforms under the rule
\begin{equation*}
g \rightarrow \tilde{g}= 2e^{\psi}g, ~~~~\text{with}~~~~\psi=\psi(x^a),
\end{equation*}
which can be expressed as
\begin{equation}
\mathcal{L}_{\xi} g_{ab} = \xi_{a;b}+ \xi_{b;a} = \psi g_{ab},\label{c1}
\end{equation}
 where $\mathcal{L}$ signifies the Lie derivative along $\xi^a$ and  $\psi(x^a)$ is
the conformal factor. In \cite{Herrera1984}, authors assumed that the vector field generating  the conformal symmetry is static and spherically symmetric within the framework of GR, which yield
\begin{equation}
\xi= \xi^0 {r} \frac{\partial}{\partial t}+\xi^1 {r} \frac{\partial}{\partial r}.\label{c2}
\end{equation}
Using this form of the conformal vector in Eq. (\ref{c1}), one obtains
\[ \xi^r \nu^\prime =\psi(r),~ \xi^t  = {\rm const.},~\xi^r  = \frac{\psi r}{2}, ~ \xi^r \lambda^\prime + 2 \xi^{r \prime} =\psi(r) .\]
These vectors are of physical significance as  (i) $\psi=0$ then Eq. (\ref{c1}) gives the Killing vector, (ii) $\psi=$ constant gives homothetic vector, and (iii) when $\psi=\psi(\textbf{x},t)$ having a conformal motion.
 
 Thus, conformal equation (\ref{c1}) for the metric (\ref{a11}) provides the following set of equations
\begin{eqnarray}
e^\nu  &=& c_2^2 r^2, \label{c3}\\ 
1-\frac{b(r)}{r}  &=& \left[\frac {\psi}{c_3}\right]^2,  \label{c4} \\ 
\xi^i &=& c_1\delta_4^i + \left[\frac{\psi r}{2}\right]\delta_r^i, \label{c5}
\end{eqnarray}
where $c_1$, $c_2$ and $c_3$ are constants of integration. Interestingly, if we rearrange the Eq. (\ref{c4}) in terms of the shape function, then the conformal factor is zero at the throat, i.e. $\psi(r_0) =0$.

\begin{figure}[t]
\centering
\includegraphics[scale=.75]{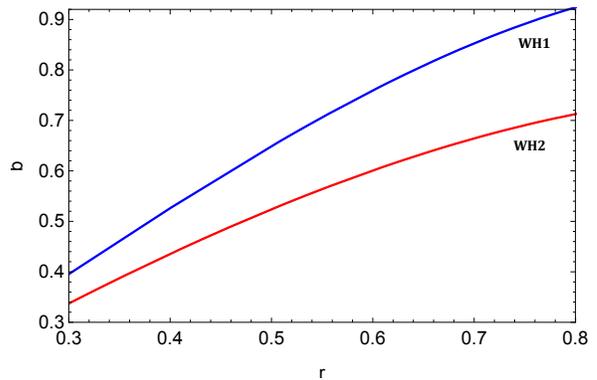}
\caption{Variation of shape functions with radial coordinate for $a=0.1,A=-2, B=0.5,\omega=-1.4,c_3=1.88$ (WH1) and $a=0.2,d=-0.014, B=0.5, n=0.14,c_3=0.13$ (WH2).}\label{f1}
\end{figure}

\begin{figure}[t]
\centering
\includegraphics[scale=.75]{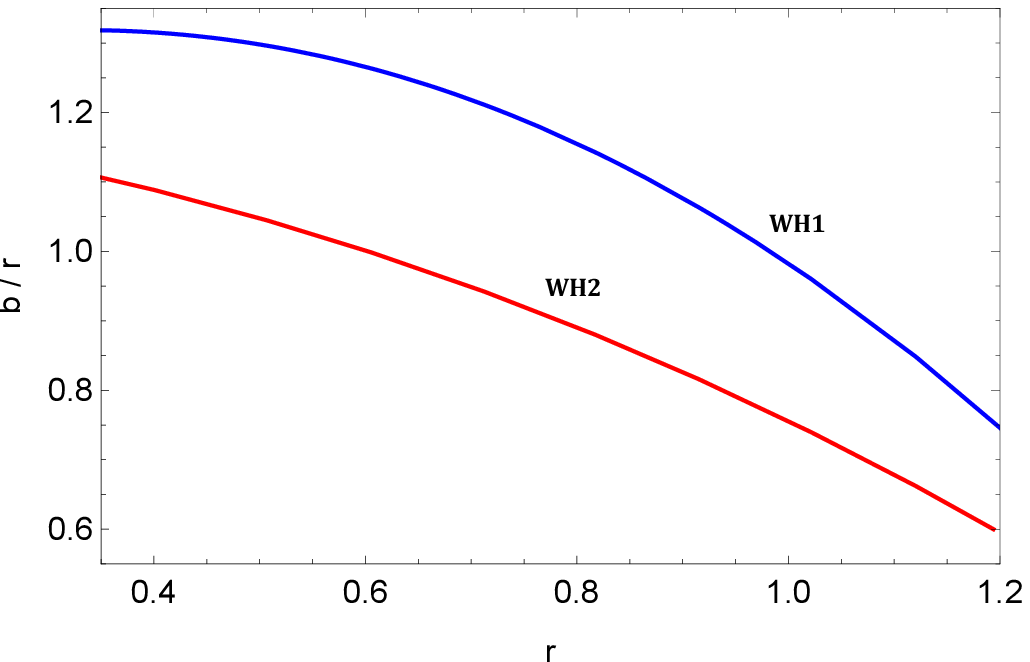}
\caption{Variation of $b/r$ with radial coordinate.}\label{f2}
\end{figure}

\begin{figure}[t]
\centering
\includegraphics[scale=.75]{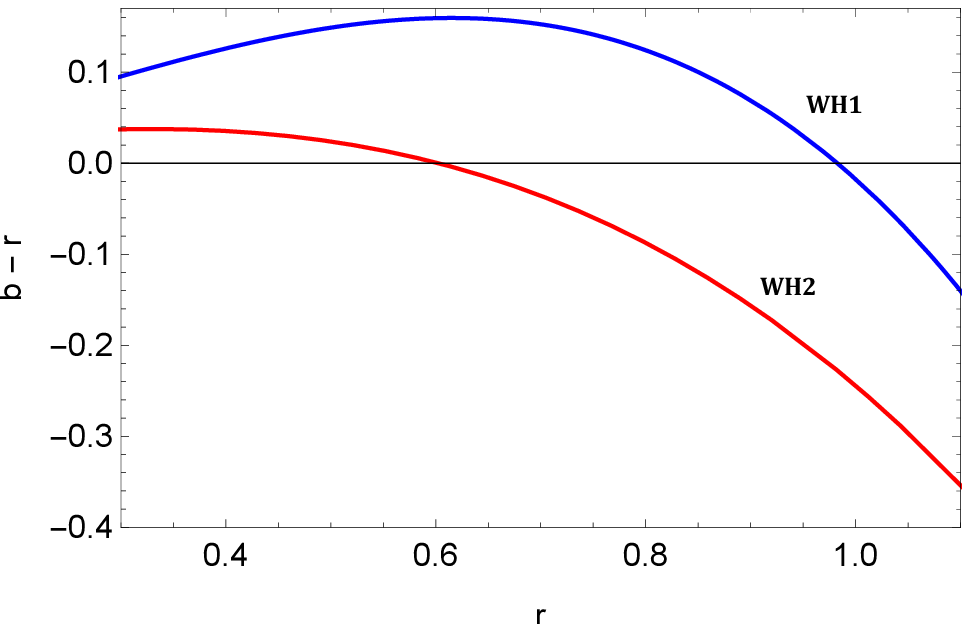}
\caption{Variation of $b-r$ with radial coordinate.}\label{f3}
\end{figure}

\begin{figure}[t]
\centering
\includegraphics[scale=.75]{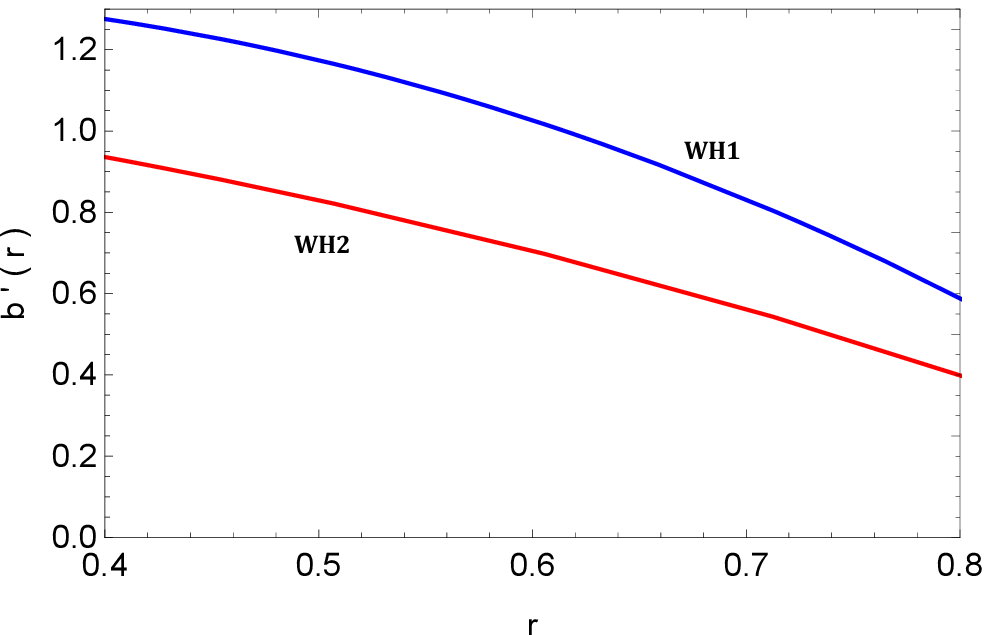}
\caption{Variation of $b'(r)$ with radial coordinate.}\label{f4}
\end{figure}

\subsection{Field equations with conformal symmetry in teleparallel gravity}

With the assumption (\ref{eq0}) the gravitational field equations describing the interior of a wormhole geometry will be imposed by the existence of conformal killing vector, so that the components of stress-energy tensor are written solely in terms of the  conformal factor. Substituting Eqs. (\ref{c3})-(\ref{c4}) into the field Eqs. (\ref{eq18})-(\ref{eq20}), we obtain
\begin{eqnarray}
4\pi \rho &=& \frac{1}{4} \left[B-\frac{2 a \psi \left(2 r \psi '+\psi \right)}{c_3^2 r^2}+\frac{2 a}{r^2}\right], \label{eq29} \\
4\pi p_r &=& \frac{1}{4} \left(\frac{6 a \psi^2}{c_3^2 r^2}-\frac{2 a}{r^2}-B\right), \label{eq30} \\
4\pi p_t &=& \frac{a \psi \left[\psi-(a+r) \psi '\right]}{2 c_3^2 r^2}-\frac{B}{4}. \label{eq31}
\end{eqnarray}

It becomes clear from Eq. (\ref{c4}) that we impose the following condition ($\psi^2)' >0$  when evaluated at the throat. The NEC asserts that for any null vector $k^{\mu}$, we have $T_{\mu\nu} k^{\mu}k^{\nu} \geq 0$. Using the Einstein field equations (\ref{eq29}) and (\ref{eq30}) provide the following relation
\begin{eqnarray}\label{eq32}
4 \pi\left(\rho+p_r\right)= \frac{a \psi  \left(\psi-r \psi '\right)}{c_3^2 r^2} .
\end{eqnarray}
The NEC at the throat is given by
\begin{eqnarray}\label{eq33}
4 \pi\left(\rho+p_r\right) |_{r_0}= - \frac{a^2}{2r^4_0}\left(1-b'(r_0)\right) < 0.
\end{eqnarray}
Taking into account the condition $b'_0 < 1$, one verifies the general condition $\left(\rho+p_r\right) |_{r_0} <0$. Therefore, the flaring-out condition entails the violation of the NEC.

\subsection{Field equations with conformal symmetry in $f(T)=aT^2+B$}

Proceeding the same, the field Eqs. (\ref{eq21})-(\ref{eq23}) are written solely in terms of the conformal factor, and the stress energy-momentum components are the following form 
\begin{eqnarray}
4\pi \rho &=& {1 \over 4} \left(B-\frac{2 a \psi \left[2 r \psi '+\psi \right]}{c_3^2 r^2}+\frac{2 a}{r^2} \right), \label{eq34}\\
4\pi p_r &=& {1 \over 4} \left(\frac{6 a \psi ^2}{c_3^2 r^2}-\frac{2 a}{r^2}-B \right), \label{eq35}\\ 
4\pi p_t &=& \frac{a \psi \left(\psi-(a+r) \psi '\right)}{2 c_3^2 r^2}-\frac{B}{4}.\label{eq36}
\end{eqnarray}
Now the NEC condition can be determine from the addition of density and pressure as
\begin{eqnarray}
4\pi (\rho+p_r) = \frac{a \psi  \left(\psi -r \psi '\right)}{ C^2_3 r^2}.
\end{eqnarray}
and the violation of NEC requires $(\psi^2)'>0$. The NEC at the throat is given by
\begin{eqnarray}\label{eq38}
4\pi (\rho+p_r)|{r_0}=\frac{a (r_0-b_0) \left(r_0 b'_0-2 b_0+r_0\right)}{r_0^4},
\end{eqnarray}

Note that the NEC, evaluated at the throat, $r_0$, is identically zero for arbitrary $r$ i.e. $(\rho+p_r)|{r_0}=0$. The same situation was found in Ref. \cite{Arellano:2006np} due to violation of flaring-out condition of the throat when axisymmetric traversable wormholes coupled to nonlinear electrodynamics. But, 
the main aim in our wormhole construction is that throughout the wormhole solution the matter obeying  the NEC or not. This needs some explanation, and we will discuss later.

\begin{figure}[t]
\centering
\includegraphics[scale=.6]{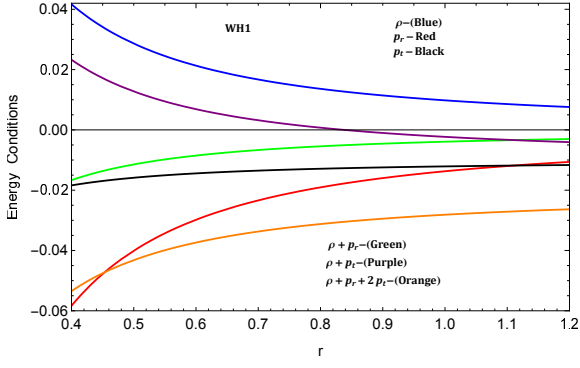}
\caption{Variation of energy conditions for WH1 with radial coordinate.}\label{f5}
\end{figure}

\begin{figure}[t]
\centering
\includegraphics[scale=.6]{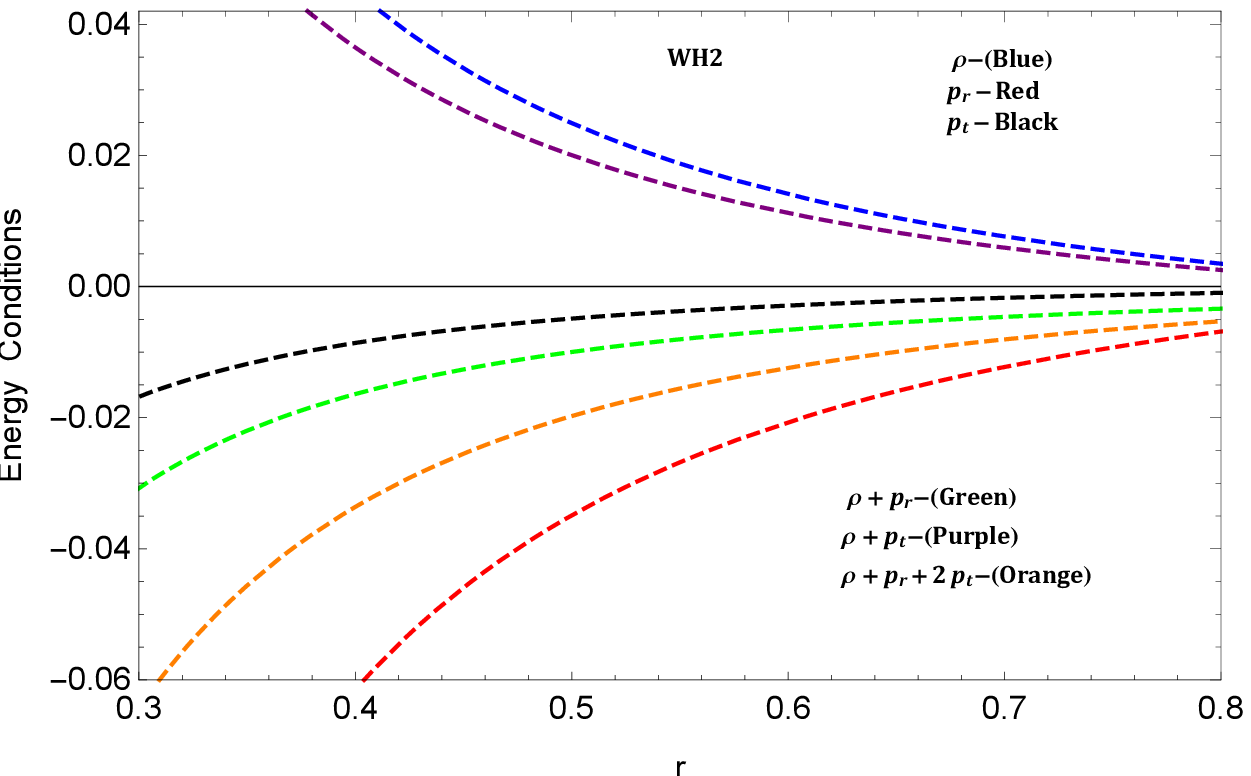}
\caption{Variation of energy conditions for WH2 with radial coordinate.}\label{f6}
\end{figure}

\begin{figure}[t]
\centering
\includegraphics[scale=.6]{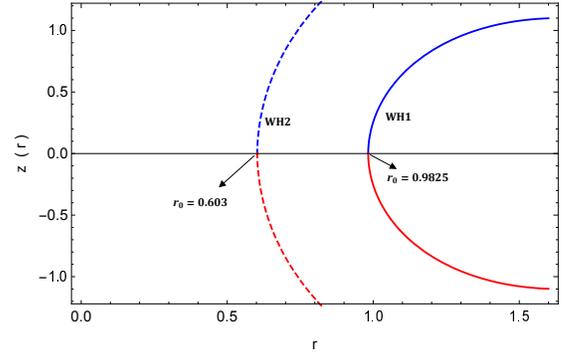}
\caption{Embedding surfaces of the two wormholes in
two dimensional space slices in $R^3$.}\label{f7}
\end{figure}

Since the solutions analyzed in this work are not asymptotically flat, so that one needs to match these interior geometries to an exterior vacuum spacetime in the asymptotic limit by taking into account thin shells, using the cut-and-paste technique \cite{Poisson:1995sv}. The appropriate framework to match the interior to the exterior solution we need the junction conditions across a timelike hypersurface for the $f(T)$ gravity. In particular, the junction conditions for $f(T)$ gravity has been performed in \cite{Velay-Vitow:2017odc} via the variational principle. It is known that the conditions for matching of two spacetime in a region given by $r > R$, if the interior metric is matched to this vacuum spacetime at $r = R > r_0$. In contradistinction for regular geometry with an interior wormhole spacetime, $r_0 \leq r \leq R$, and a Schwarzschild–like vacuum, $r_h < R \leq r < \infty$, at a junction interface. In addition, the junction conditions in this theory have
been worked out \cite{delaCruz-Dombriz:2014zaa}.

\section{TEGR Wormholes }\label{sec6}

For the simplest linear function $f(T) =aT+b$, we will discuss some particular wormhole solutions. To start with, we shall first consider the isotropic case and then various choices for the form function.

\subsection{Isotropic wormhole solution}
First, we shall restrict our investigation in isotropic condition i.e. $p_r=p_t$, which on imposing to the field equations (\ref{eq30}) and (\ref{eq31}), one can get the following solution:
\begin{equation}\label{eq39}
\psi = \frac{\sqrt{c_3^2 (a+r)^4+e}}{\sqrt{2} (a+r)^2},
\end{equation}
thus substituting this value into (\ref{c4}), we have 
\begin{equation}\label{eq40}
b(r) = \frac{r}{2} \left[1-\frac{e}{c_3^2 (a+r)^4}\right],
\end{equation}
where $e$ is the constant of integration. 

It becomes clear that this form of the shape function gives non asymptotically flat geometries, i.e. $b(r)/r \rightarrow 1/2$ as $r \rightarrow \infty$ and $b'(r_0)  \nless 1$. This implies that there are no wormhole solutions sustained for isotropic pressure in TEGR.

\subsection{Wormhole (WH1) solution with $p_r = \omega \rho$: }

To solve the field equations (\ref{eq29})-(\ref{eq31}), we assume an additional information as a linear equation of state (EoS) $p_r =\omega \rho$, and the corresponding solutions can be written as 
\begin{eqnarray}\label{eq46}
\psi(r) &=&  \frac{1}{\sqrt{6 a (\omega +3)}} \bigg[ 6 a A (\omega +3) r^{-\frac{\omega +3}{\omega }}+c_3^2 \big\{6 a (\omega +1) \nonumber \\
&& +B r^2 (\omega +3)\big\} \bigg]^{1/2}, \label{eq40}
\end{eqnarray}
where $A$ is an integrating constant. Note that the solution given by (\ref{eq40}) holds for $\omega$ $\neq -3$  only.

Immediately, one can write down the corresponding shape function as
\begin{eqnarray}\label{eq47}
b(r) &=& \frac{2 r}{\omega +3}-\frac{B r^3}{6 a}-\frac{A r^{-3/\omega }}{c_3^2}. \label{eq41}
\end{eqnarray}
It is clear that the solutions are asymptotically flat, i.e. $b(r)/r \rightarrow 0$ as $r \rightarrow \infty$. Note that from Eq. (\ref{eq41}) when $\omega =0$, one obtains $b(r) = \frac{2}{3}r-\frac{Br^3}{a}$ which presents a non-asymptotically flat wormhole geometry.  Further, to be a wormhole solution,  we deduce $b'(r)$ =$\frac{2}{\omega +3}-\frac{B r^2}{2 a}+\frac{3 A r^{-3/\omega-1}}{C_3^2 \omega }$, and impose the the condition $b'(r_0)< 1$.  In fact, we fix the range of parameters from the criterion.

We now discuss the possibility of sustaining a traversable wormhole in spacetime via exotic matter made out of phantom energy EoS, $p_r =\omega \rho$ with $\omega < -1$, and the stress-energy tensor components are given by 
\begin{eqnarray}\label{eq44}
\rho &=&  \frac{a }{8 \pi  r^3} \left(\frac{3 A r^{-3/\omega }}{c_3^2 \omega }+\frac{2 r}{\omega +3}\right) ,\label{eq45}\\
\rho+p_r &=& \frac{a (\omega +1) }{8 \pi  r^3} \left(\frac{3 A r^{-3/\omega }}{c_3^2 \omega }+\frac{2 r}{\omega +3}\right), \\
\rho+p_t &=& \frac{1}{48 \pi  r^4} \bigg[\frac{3 a A (\omega +3) (a+3 r) r^{-3/\omega }}{c_3^2 \omega }+a r^2 \nonumber \\
&& (6-B r)-3 B r^4 \bigg],\\
\rho+p_r+2p_t &=& \frac{1}{24 \pi  r^4} \bigg[\frac{3 a A r^{-3/\omega } (a (\omega +3)+6 r (\omega +1))}{c_3^2 \omega } \nonumber \\
&& +r^2 \left(\frac{12 a (\omega +1)}{\omega +3}-a B r-3 B r^2\right) \bigg].
\end{eqnarray}

To check the NEC evaluated at the throat is given by 
\begin{equation}\label{eq49}
(\rho+p_r)|_{r_0} = \frac{a (\omega +1) }{8 \pi  r^3_0} \left(\frac{3 A r^{-3/\omega }}{c_3^2 \omega }+\frac{2 r_0}{\omega +3}\right).
\end{equation}
 Clearly, in this case for $\omega \neq -1$ and $\omega \neq -3$, we consider the interval $-3 < \omega < -1$, implying the violation of the NEC at the throat i.e. the throat needs to open with phantom energy. 

In Figs. \ref{f1}-\ref{f4}, we plot $b(r)$, $b(r)/r$, $b(r)-r$ and $b'(r)$ respectively, for $a=0.1,A=-2, B=0.5,\omega=-1.4, c_3=1.88$ (WH1). Note that, $b(r)-r$ cuts $r$-axis at $r_0=0.9825$ corresponds to the throat radius of WH1.
This situation is shown graphically in Fig. \ref{f7}. In fact, wormhole  geometries fulfilling the required condition $b'(0.9825)\approx 0.843 <1$, that can see directly from Fig. \ref{f4}.

On the other hand, Fig. \ref{f5} depicts the behavior of the energy conditions using the field Eqs. (\ref{eq44})-(\ref{eq46}). One may also see that a matter content with a radial pressure having a phantom EoS i.e $\omega < -1$ and everywhere positive energy density $\rho >0$. Note that it will be valid far from the throat. More specifically, NEC is violated due to $\rho+p_r < 0$. The embedded surface can be evaluated from Eq. (\ref{emb}). The Eq. (\ref{emb})
is strongly dependent on numerical values of the
model parameters and thereby adopting the numerical approach using ``Mathematica" command we plot the embedding surface Fig. \ref{f7}.

Since the redshift function $\nu(r)$ does not tend to zero when $r \rightarrow \infty $ due to the conformal symmetry. Thus, a constant limit for $\nu(r)$  would also allow us to obtain asymptotically at solutions under time reparametrization. Leading to Ref. \cite{Boehmer:2007md}, the conditions for matching of two spacetime in a region given by $r > R$, if the interior metric is matched to the vacuum spacetime at $r = R > r_0$. Several examples of conformal symmetry have been found mainly in wormhole physics \cite{Rahaman:2013ywa,Rahaman:2014dpa,Kuhfittig:2015cea,Kuhfittig:2015xwa}.

\subsection{Wormhole (WH2) solution with $p_t = np_t$:}
Another closed-form solution is derived by taking $ p_t=n p_r$ (see Ref. \cite{Rahaman:2006xa,Moraes:2017mir} for more details) and then deduce the following relationship:

\begin{eqnarray}
\psi (r) &=& (a+b) \bigg[ \frac{c_3^2}{a} \bigg(\frac{a^2 B (n-1)+2a n}{2 (3 n-1) (a+r)^2}-\frac{2 a B (n-1)}{(6 n-1) (a+r)} \nonumber \\
&& +\frac{B (n-1)}{6 n}\bigg)+d (a+r)^{-6 n} \bigg]^{1/2}, \label{eq48}
\end{eqnarray}
where the state parameter $n$ is a constant. We obtain from Eq. (\ref{eq48}) yielding for the shape function 
\begin{eqnarray}
b(r) &=& \frac{r}{6} \bigg[-\frac{B (n-1) (a+r)^2}{a n}+\frac{12 B (n-1) (a+r)}{6 n-1}- \nonumber \\
&& \frac{3 [a B (n-1)+2 n]}{3 n-1}-\frac{6 d (a+r)^{2-6 n}}{c_3^2}+6 \bigg]. \label{eq47}
\end{eqnarray}
In this case for $n <1$ implies $b(r)/r \rightarrow 0$ as $r \rightarrow \infty$ \i.e. asymptotically flat spacetimes. 

Using the Einstein field equations (\ref{eq29})-(\ref{eq31}) with the following shape function, one obtains
\begin{eqnarray}
&& \rho = \frac{1}{48 \pi  r^2} \bigg[\frac{f_1(r)/(3 n-1)}{n (6 n-1)}-\frac{6 a d (a-6 n r+3 r)}{c_3^2 (a+r)^{6 n-1}} \bigg], \label{eq54}\\
&& \rho+p_r = \frac{a}{24 \pi  r^2} \bigg[\frac{6 d (a+3 n r) (a+r)^{1-6 n}}{c_3^2}+ \nonumber \\
&&  \frac{a B (n-1)-B (n-1) (3 n-1) r+6 (6 n-1) n^2}{n (3 n-1) (6 n-1)}\bigg], \\
&& \rho+p_t = \frac{(a+r)^{-6 n}}{48 \pi  c_3^2 n (6 n-1) r^2} \bigg[c_3^2 (a+r)^{6 n} \Big\{a^2 B (n-1)- \nonumber \\
&& 2 a B (n-1) (3 n-2) r+6 a n (6 n-1)-3 B \nonumber \\
&& (n-1) (6 n-1) r^2\Big\}+6 a d n \{9 n (2 n-1)+1\} \nonumber \\
&& (a+3 r) (a+r) \bigg]. \\
&& \rho+p_r+2p_t = \frac{(a+r)^{-6 n}}{24 \pi  c_3^2 n (3 n-1) (6 n-1) r^2} \bigg[c_3^2 f_3(r) (a+r)^{6 n} \nonumber \\
&& +6 a d n \{9 n (2 n-1)+1\} (a+r) (3 a n+a+6 n r) \bigg].
\end{eqnarray}

Note we have used the notations
\begin{eqnarray}
f_1(r) &=& a^2 (B-B n)+4 a B (n-1) (3 n-1) r+6 a n \nonumber \\
&& (2 n-1) (6 n-1)+3 B (9 n (2 n-1)+1) r^2 \nonumber \\
f_2(r) &=& a^2 B (n-1)-2 a [n \{B (3 n-4) r-6 n+1\}+B r]  \nonumber \\
&& +B \{9 (1-2 n) n-1\} r^2 \nonumber\\
f_3(r) &=& a^2 B (n-1) (3 n+1)-a B (n-1) (3 n-1) (6 n+1) \nonumber\\
&& r+12 a n^2 (6 n-1)-3 B n \{9 n (2 n-1)+1\} r^2. \nonumber
\end{eqnarray}

The above expression must be investigated at the throat $r_0$ to check the NEC, which is given by
\begin{eqnarray}
(\rho+p_r)|_{r_0} &=& \frac{a}{24 \pi  r_0^2} \bigg[\frac{6 d (a+3 n r_0) (a+r_0)^{1-6 n}}{c_3^2}+ \nonumber \\
&& \hspace{-2 cm} \frac{a B (n-1)-B (n-1) (3 n-1) r_0+6 (6 n-1) n^2}{n (3 n-1) (6 n-1)}\bigg].
\end{eqnarray} 
148

We plot the quantities $b(r)$, $b(r)/r$, $b(r)-r$ and $b'(r)$ in  Figs. \ref{f1}-\ref{f4}.  For the figures we  consider $a=0.2,d=-0.014, B=0.5, n=0.14,c_3=0.13$ (WH2). We can see from Fig. \ref{f3} and  Fig. \ref{f7} that $b(r)-r$ cuts $r$-axis at $r_0 = 0.603$, which is the throat of WH2. One verifies, $b'(0.603) \approx 0.534 <1$ is shown in Fig. \ref{f4}. 
 
Let us emphasize again the energy conditions. This situation differs from the above discussion  that $\rho+p_r$ and $\rho+p_t$ both are negative, whereas $\rho$ is positive throughout the spacetime. As shown in Fig. \ref{f6}, for some fixed parametric values,  the NEC is violated in a small region around $r_0$. The embedding surface $z(r)$ in 3-D Euclidean space can be obtained from Eq. (\ref{emb}). In Fig. \ref{f7} we show the wormhole embedding diagrams for the values of WH2. However, for this solution as well it is not integrable and therefore adopting the numerical approach using ``NIntegrate" command within the limits $r_0\le r \le R$. It becomes clear that all embedding surfaces flare outward.

\begin{figure}[t]
\centering
\includegraphics[scale=.75]{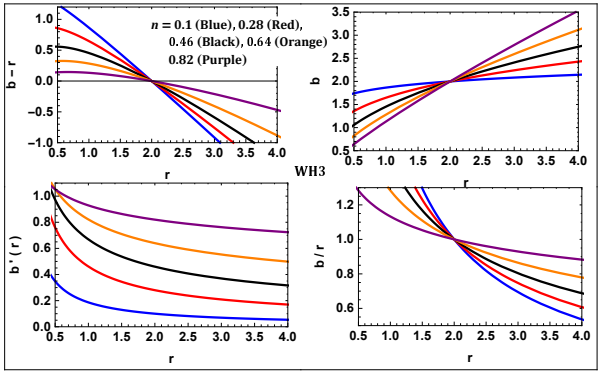}
\caption{Characteristics of the shape function of WH3 for $a=0.6, R=2, B=0.5, c_3=1.165$.}\label{f8}
\end{figure}

\begin{figure}[t]
\centering
\includegraphics[scale=.8]{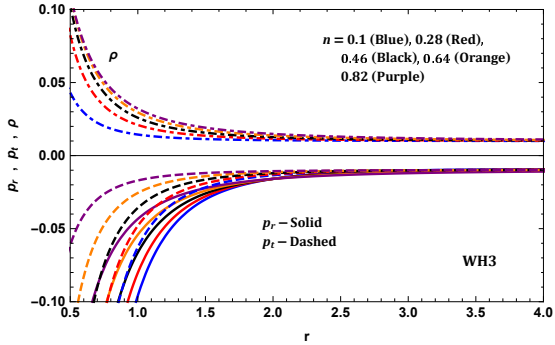}
\caption{Trends of density and pressures for WH3.}\label{f9}
\end{figure}

\begin{figure}[t]
\centering
\includegraphics[scale=.8]{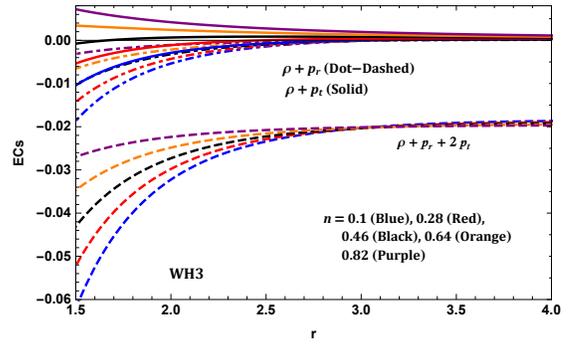}
\caption{Variations of energy conditions for WH3.}\label{f10}
\end{figure}

\subsection{Wormhole (WH3) solution with $b(r)= r_0 \left(r/r_0\right)^n$:}

To proceed further we ansatz the shape function as $b(r)=r_0 \left(r/r_0\right)^n$. In this case, we need to impose $0< n <1$, and corresponding  conformal factor takes the form 
\begin{eqnarray}
\psi = c_3 \sqrt{1-\frac{r_0}{r}  \left({ r \over r_0}\right)^n}.
\end{eqnarray}
Our aim here is to explore the local energy conditions, and we examine the WEC. Taking into account the diagonal energy momentum tensor, the the WEC implies $\rho \geq 0$,  $\rho+p_r \geq 0$ and $\rho+p_t \geq 0$. Note that last two inequalities reduce to the null energy condition (NEC). The components of the energy-momentum tensor (\ref{eq29})-(\ref{eq31}) then take the form 
\begin{eqnarray}
\rho &=& \frac{2 a n r_0 \left(r/ r_0\right)^n+B r^3}{16 \pi  r^3},  \label{eq53} \\
\rho+p_r &=& \frac{a \left[(n-3) r_0 \left(\frac{r}{r_0}\right)^n+2 r\right]}{8 \pi  r^3}, \label{eq54}\\
\rho+p_t &=& \frac{a (n-1) r_0 (a+3 r), \left(\frac{r}{r_0}\right)^n+2 a r^2}{16 \pi  r^4} \label{eq55}
\end{eqnarray}
\begin{eqnarray}
\rho+p_r+2p_t &=& \frac{1}{8 \pi  r^4} \Big[a r_0 [a (n-1)+2 (n-3) r] \nonumber \\
&& \left(\frac{r}{r_0}\right)^n+4 a r^2-B r^4 \Big]\label{eq56}.
\end{eqnarray}
Note that at the throat, Eq. (\ref{eq54}) reduces to 
\begin{eqnarray}
(\rho+p_r) |_{r_0} = \frac{a (n-1)}{8 \pi  r_0^2}.
\end{eqnarray}
Taking into account the condition $b' ({r_0}) < 1$, and for $0<n < 1$, one may verify that $(\rho+p_r) |_{r_0}< 0$. Fig. \ref{f8} shows the behaviour of $b(r)$, $b(r)-r$, $b'(r)$ and $b(r)/r$, respectively. In this case,  $b(r)-r$ cuts the $r$-axis at $r_0= 2$, which is the throat radius for WH3 (see Fig. \ref{f13}).  

From the graphical behavior of the energy conditions in terms of (\ref{eq53})-(\ref{eq56}) are presented in Fig. \ref{f9}-\ref{f10}.  From Fig. \ref{f9} , we see that the energy density is positive throughout the whole spacetime, while the radial and transverse pressures are negative, and both tend to zero in the asymptotic limit by construction. Moreover, we observe that for fixed values of the parameters $a=0.6, R=2, B=0.5, c_3=1.165$ the NEC is violated due $\rho+p_r <0$; note also that $\rho >0$ for different values of $n$. The embedding surface $z(r)$ in 3-D Euclidean space obtained by using Eq. (\ref{emb}) through ``NIntegrate" in Mathematica and  shown in Fig. \ref{f13}.

\begin{figure}[t]
\centering
\includegraphics[scale=.8]{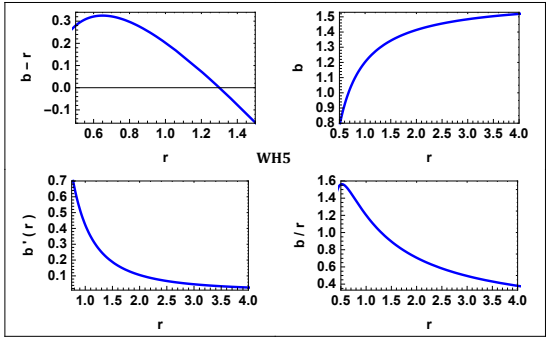}
\caption{Characteristics of the shape function of  WH5 for $a=0.2, R=1.3, \alpha=0.25, B=0.3, c_3=1.165$.}\label{f11}
\end{figure}

\begin{figure}[t]
\centering
\includegraphics[scale=.8]{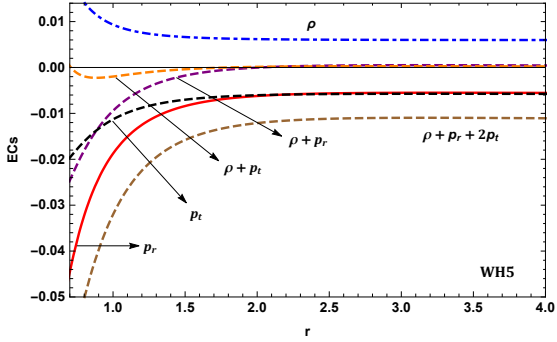}
\caption{Trends of energy conditions for WH5.}\label{f12}
\end{figure}

\subsection{Wormhole (WH5) solution with $b(r) = \alpha \left(1-r_0/r \right)+r_0$:}

Now, we will consider the wormhole solution generated by imposing the shape function in the form 
$b(r) = \alpha \left(1-r_0/r \right)+r_0 $. For this specific case, $b'( {r_0})=\alpha/r_0 < 1$ when $\alpha < 1$, and
Eq. (\ref{c4}) provides the following solution
\begin{equation}
\psi = c_3 \sqrt{1-\frac{\alpha  R \left(1-\frac{R}{r}\right)+R}{r}}.
\end{equation}
In this case the tress energy components are given by
\begin{eqnarray}
\rho &=& \frac{2 a \alpha  r_0^2+B r^4}{16 \pi  r^4}, \\
\rho+p_r &=& \frac{a \left(2 r^2-3 (\alpha +1) r r_0+4 \alpha  r_0^2\right)}{8 \pi  r^4},\\
\rho+p_t &=& \frac{a}{16 \pi  r^5} \Big[2 r^3-\alpha  r_0 (a+3 r) (r-2 r_0) \nonumber \\
&& -r r_0 (a+3 r)\Big],\\
\rho+p_r+2p_t &=& -\frac{1}{8 \pi  r^5} \Big[a^2 r_0 (\alpha  r+r-2 \alpha  r_0)+2 a r \big\{-2 r^2 \nonumber \\
&& +3 (\alpha +1) r r_0-4 \alpha  r_0^2\big\}+B r^5 \Big].
\end{eqnarray}
We now check the energy condition at the throat of the wormhole, which is
\begin{eqnarray}
(\rho+p_r)|_{r_0} = \frac{a (\alpha-1)}{8 \pi r{_0}^2},
\end{eqnarray}

\begin{figure}[t]
\centering
\includegraphics[scale=.8]{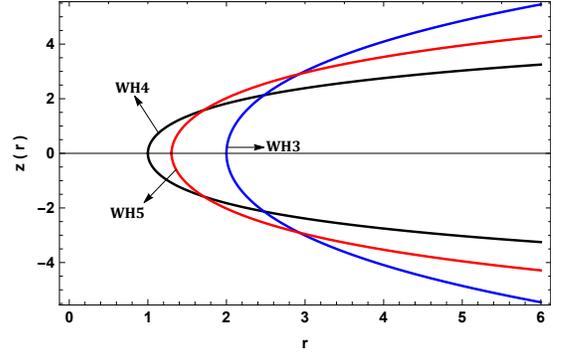}
\caption{The figure shows three dimensional wormhole embedding diagrams for WH3, WH4 and WH5.}\label{f13}
\end{figure}

One can easily check that for $0<\alpha < 1$ the condition $(\rho+p_r) |_{r_0}< 0$.  Fig. \ref{f11} depicts $b(r)$, $b(r)-r$, $b'(r)$ and $b(r)/r$, in terms of $r$ for  $a=0.2, r_0=1.3, \alpha=0.25, B=0.3$ and $c_3=1.165$, respectively. It can be noted that $b(r)$ cuts the $r$-axis at $r_0= 1.3$ for WH5 (see Fig. \ref{f13}). Moreover, from the  Fig. \ref{f12}, we see that $\rho+p_r < 0$ and $\rho+p_t <0$, while $\rho >0$ throughout the spacetime lead to the violation of WEC, and consequently NEC also. The embedding surface can be determine using (\ref{emb}), and found to
\begin{eqnarray}
z(r) &=& 2 r_0 \bigg[(\alpha +1) \log \left(\sqrt{r-\alpha  r_0}+\sqrt{r-r_0}\right)-\sqrt{\alpha } \nonumber \\
&& \tanh ^{-1}\left(\frac{\sqrt{\alpha } \sqrt{r-r_0}}{\sqrt{r-\alpha  r_0}}\right)\bigg].
\end{eqnarray}
The embedding surface are shown in Fig. \ref{f13}. 
For a full visualization of the surface sweep through a $2\pi$ rotation around the $z$−axis, as depicted in Fig. \ref{f23}.

\section{$f(T)$ Wormholes } \label{sec7}

In this section we proceed in an attempt to analytically solve the basic equations by considering a power-law $f(T)=aT^2+B$ function, and for the anisotropic fluid. Here we assume different shape function in finding wormhole solutions. Since solving the differential equations (\ref{eq34}-\ref{eq36}) in general, are too complicated for the choices of $p_r=\omega \rho$ or $p_t=n p_r$. Therefore, in order to simplify the analysis, we will consider restrictions on the choice of shape functions.

\begin{figure}[t]
\centering
\includegraphics[scale=.7]{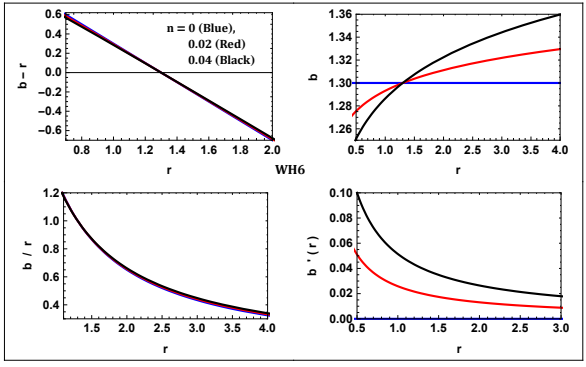}
\caption{Characteristics of shape functions for WH6 with $a=0.5, R=1.3, c_3=1.165$ and $B=0.2$.}\label{f14}
\end{figure}

\begin{figure}[t]
\centering
\includegraphics[scale=.8]{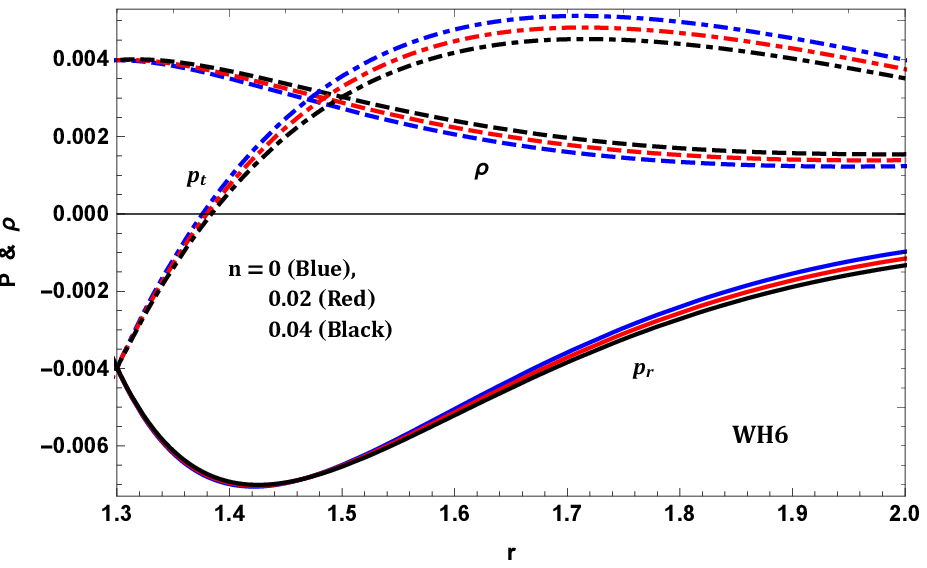}
\caption{Variations of density and pressures for WH6.}\label{f15}
\end{figure}
\begin{figure}[t]
\centering
\includegraphics[scale=.8]{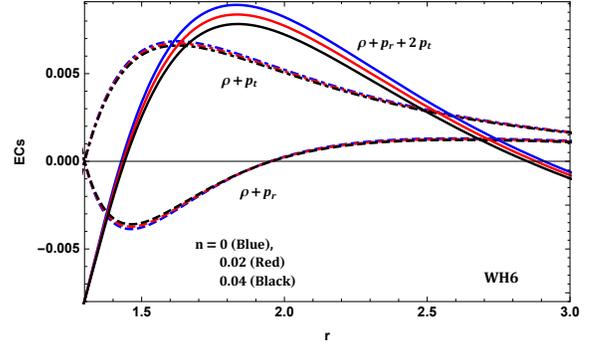}
\caption{Variations of energy conditions for WH6.}\label{f16}
\end{figure}

\subsection{Wormhole (WH6) with $b(r) = r_0 \left(r/r_0\right)^n$:}

Here, we assume the simplest and viable power-law form of the $f(T)=aT^2+B$ model with $a$ and $B$ are constants. Consider the specific shape function given by $b(r)= r_0 \left(r/r_0\right)^n$, we obtain the conformal factor as follows:
\begin{eqnarray}
\psi(r) &=& c_3 \sqrt{1-\frac{r_0 \left(r/r_0\right)^n}{r}},
\end{eqnarray}
Inserting these functions into the stress-energy tensor
profile, Eqs. (\ref{eq34})-(\ref{eq36}), provides the following expressions
\begin{eqnarray}
\rho &=& \frac{B r^6-12 a }{16 \pi  r^6}  \left[3 r-(2 n+3) r_0 \left(\frac{r}{r_0}\right)^n\right] \nonumber \\
&& \left[r-r_0 \left(\frac{r}{r_0}\right)^n\right], \\
\rho + p_r &=& \frac{3 a }{2 \pi  r^6}  \left[(n-3) r_0 \left(\frac{r}{r_0}\right)^n+2 r\right] \nonumber \\
&& \left[r-r_0\left(\frac{r}{r_0}\right)^n\right],  \\
\rho + p_t &=& \frac{3 a \left[r-r_0 \left(\frac{r}{r_0}\right)^n\right]}{2 \pi  r^5},\\
\rho + p_r + 2p_t &=& \frac{1}{8 \pi  r^6} \Big[12 a \left\{r-r_0 \left(\frac{r}{r_0}\right)^n\right\} \nonumber \\
&& \Big\{7 r-(n+6) r_0 \left(\frac{r}{r_0}\right)^n\Big\}-B r^6 \Big].
\end{eqnarray}

The geometrical properties and characteristics for these shape function is depicted in Fig. \ref{f14}. In this case, $b(r)- r$ cuts the $r$-axis at $r_0 = 1.3$, which is the throat radius for WH6 when $n <1$ (see Fig. \ref{f23} also).

Let us focus the case when $n<1$ to visualise better the behaviour of the energy conditions. Indeed, one may see that $(\rho+p_r)|_{r_0}=0$ at the throat of the wormhole (as mentioned in Sec V.2), however Fig. \ref{f15} shows the validity of $\rho> 0$. Thus, one can in principle construct wormhole solutions that satisfy the NEC at the wormhole throat. 

In Figs. \ref{f14} and \ref{f15} we plot the quantities $\rho$, $\rho+p_r$, $\rho+p_t$ and $\rho+p_r+2p_t$. For the figures we have considered $a=0.5, r_0=1.3, c_3=1.165$ and $B=0.2$, respectively. For these choices, the quantities outside the throat $\rho+p_r<0$ but $\rho+p_t >0$, which in principle violation of the NEC implying that the WEC is also violated. Therefore, the range of $0<r< r_0$, the solution obeys the NEC, whereas for any value of $r>r_0$, violation of the NEC outside the throat radius and goes up to the radius $R$ as we need to match at some $ r \le R$, so that we can use the junction conditions to match the interior wormhole solution to the exterior vacuum spherically symmetric solution for finite redshit function. The embedding surface $z(r)$ is calculated numerically in Mathematica using ``NIntegrate" package for $r\ge r_0$, and shown in Fig. \ref{f23}.

\subsection{Wormhole (WH7) with $b(r) = \alpha  r_0^3 \log \left(r_0 /r\right)+ r_0$:}

Consider the specific shape function  $b(r)=\alpha  r_0^3 \log \left(r_0/r \right)+r_0$. For this choice, the conformal factor becomes
\begin{eqnarray}
\psi &=& c_3 \sqrt{1-\frac{\alpha  r_0^3 \log \left(r_0/r\right)+r_0}{r}},
\end{eqnarray}
and corresponding the stress energy components are

\begin{figure}[t]
\centering
\includegraphics[scale=.7]{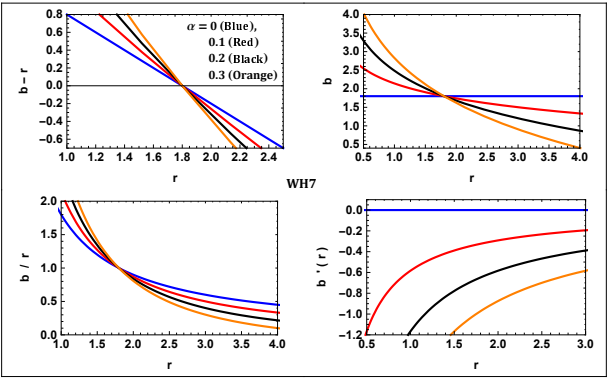}
\caption{Characteristics of shape function for WH7 with $a=0.5, R=1.8, c_3=1.165, B=0.2$.}\label{f17}
\end{figure}

\begin{figure}[t]
\centering
\includegraphics[scale=.8]{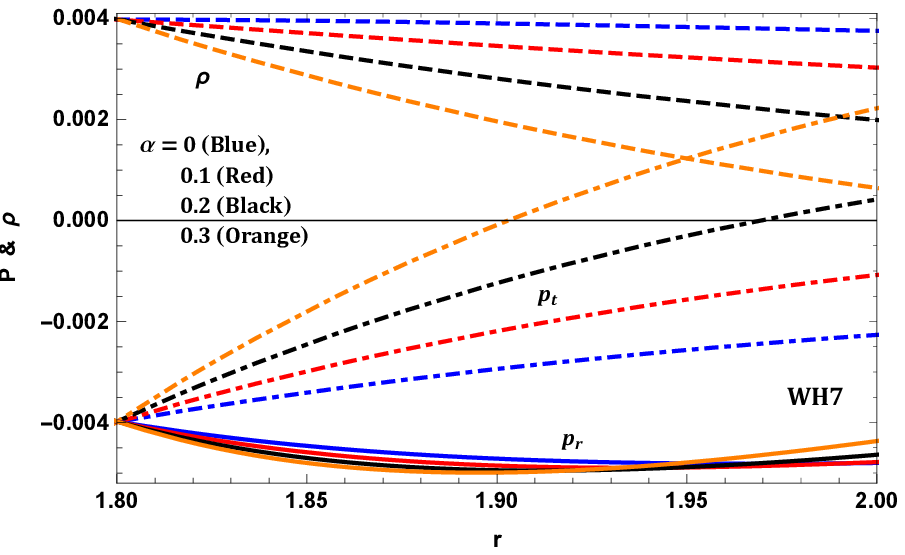}
\caption{Variation of density and pressures for WH7.}\label{f18}
\end{figure}

\begin{figure}[t]
\centering
\includegraphics[scale=.8]{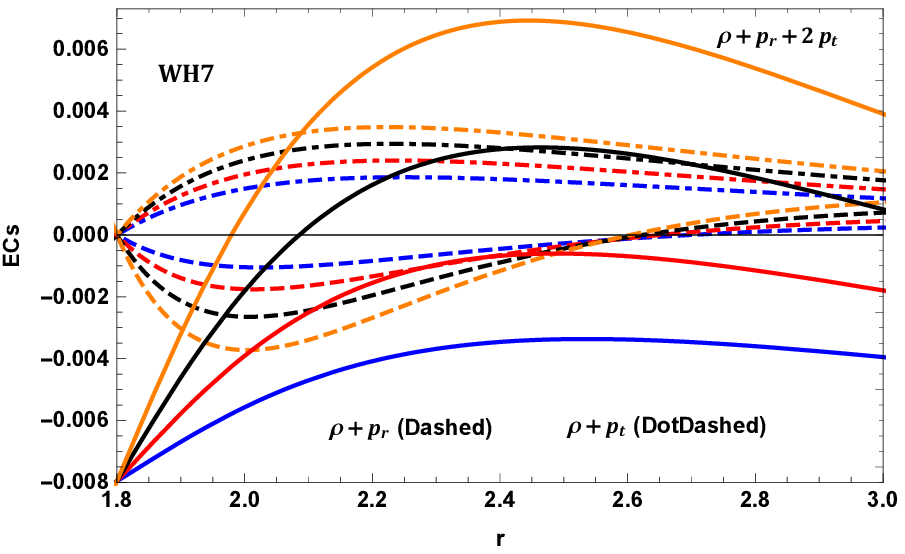}
\caption{Variation of energy conditions for WH7.}\label{f19}
\end{figure}

\begin{eqnarray}
&&\rho = \frac{1}{16 \pi  r^6} \Big[B r^6-36 a \alpha ^2 r_0^6 \log ^2\left(\frac{r_0}{r}\right)-12 a (r-r_0) (3 r \nonumber \\
&&  +2 \alpha  r_0^3-3 r_0)+24 a \alpha  r_0^3 \left(3 r+\alpha  r_0^3-3 r_0\right) \log \left(\frac{r_0}{r}\right) \Big]\\
&& \rho + p_r = \frac{3 a}{2 \pi  r^6} \bigg[3 \alpha ^2 r_0^6 \log ^2\left(\frac{r_0}{r}\right)+\alpha  r_0^3 (6r_0-5 r+\alpha  r_0^3) \nonumber \\
&&  \log \left(\frac{r_0}{r}\right)+(r-r_0) \big\{2 r-r_0 \left(\alpha  r_0^2+3\right)\big\}\bigg]
\end{eqnarray}
\begin{eqnarray}
&& \rho + p_t = \frac{3 a \left(r-\alpha  r_0^3 \log \left(\frac{r_0}{r}\right)-r_0\right)}{2 \pi  r^5}, \\
&& \rho + p_r + 2p_t = \frac{1}{8 \pi  r^6} \bigg[72 a \alpha ^2 r_0^6 \log ^2\left(\frac{r_0}{r}\right)+12 a (r-r_0) \nonumber \\
&& \left(7 r+\alpha  r_0^3-6 r_0\right)-12 a \alpha  r_0^3 (13 r+\alpha  r_0^3-12 r_0) \log \left(\frac{r_0}{r}\right) \nonumber \\
&& -B r^6 \bigg].
\end{eqnarray}
In light of the flaring-out condition at the wormhole throat $b'_{0}<1$, the parameters $\alpha $ have to meet the requirement $\alpha <1$. With $a=0.5, R=1.8, c_3=1.165$, and $B=0.2$, we plot $b(r)$, $b(r)-r$, $b'(r)$ and $b(r)/r$ in Fig. \ref{f17}.  In addition to this, the wormhole throat located at $r_0=1.8$. 

Here also the situation is same as in the previous discussion i.e. at the throat $(\rho+p_r)|_{r_0}=0$ and $(\rho+p_t)|_{r_0}=0$,  however $\rho> 0$ inside and outside the throat for the specific value of  $a=0.5, R=1.8, c_3=1.165$, and $B=0.2$ with different values of $\alpha$. Moreover, in Fig. \ref{f18}, we  plot $\rho$, $p_r$ and  $p_t$, while in Fig. \ref{f19} the behavior of the energy conditions are shown outside the throat radius. The fundamental wormhole conditions, namely, NEC is violated outside the throat, as can be readily verified from Fig. \ref{f18}. We explore the geometrical properties of these solutions via the embedding diagram  which is calculated numerically in Mathematica using ``NIntegrate" package for $r\ge r_0$ and shown in Fig. \ref{f23}.

\subsection{Wormhole (WH8) with $b(r) = \alpha  r_0 \left(1-r_0/r\right)+r_0$:}

We consider wormhole with the following shape function $b(r)=\alpha  r_0 \left(1-r_0/r\right)+r_0$. Thus, the conformal factor becomes
\begin{eqnarray}\label{eq79}
\psi &=& c_3 \sqrt{1-\frac{\alpha  r_0 \left(1-r_0/r\right)+r_0}{r}},
\end{eqnarray}

\begin{figure}[t]
\centering
\includegraphics[scale=.75]{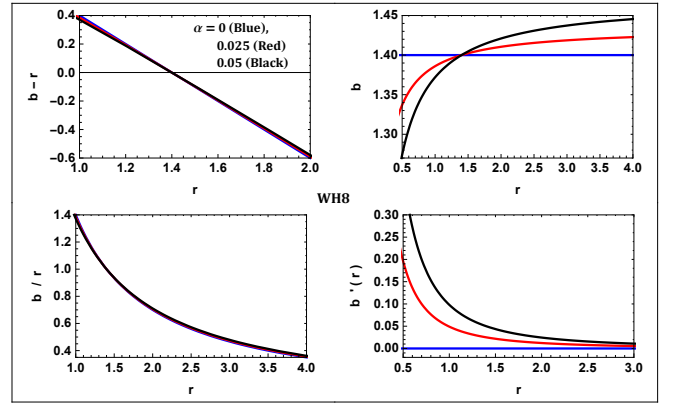}
\caption{Characteristics of shape function for WH8 with $a=0.5, R=1.4, c_3=1.165, B=0.2$.}\label{f20}
\end{figure}

\begin{figure}[t]
\centering
\includegraphics[scale=.75]{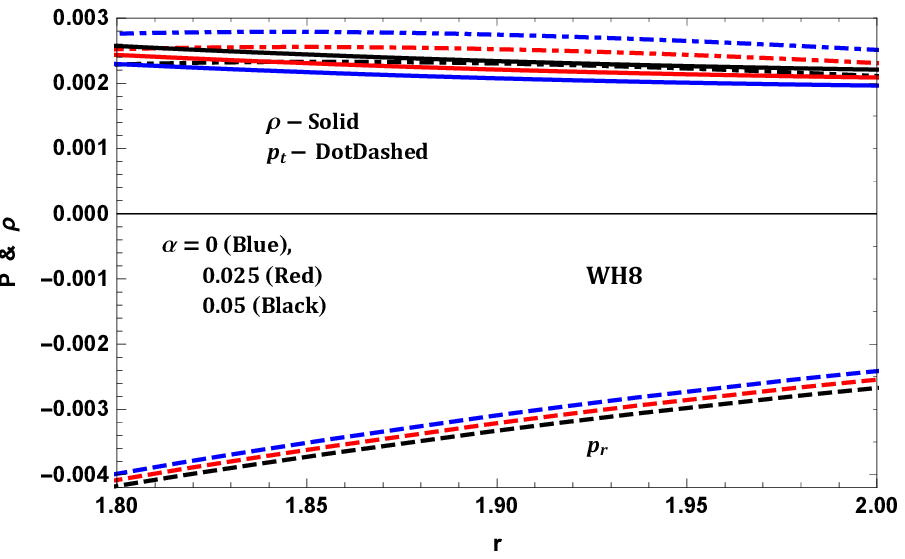}
\caption{Variation of density and pressures for WH8.}\label{f21}
\end{figure}

\begin{figure}[t]
\centering
\includegraphics[scale=.8]{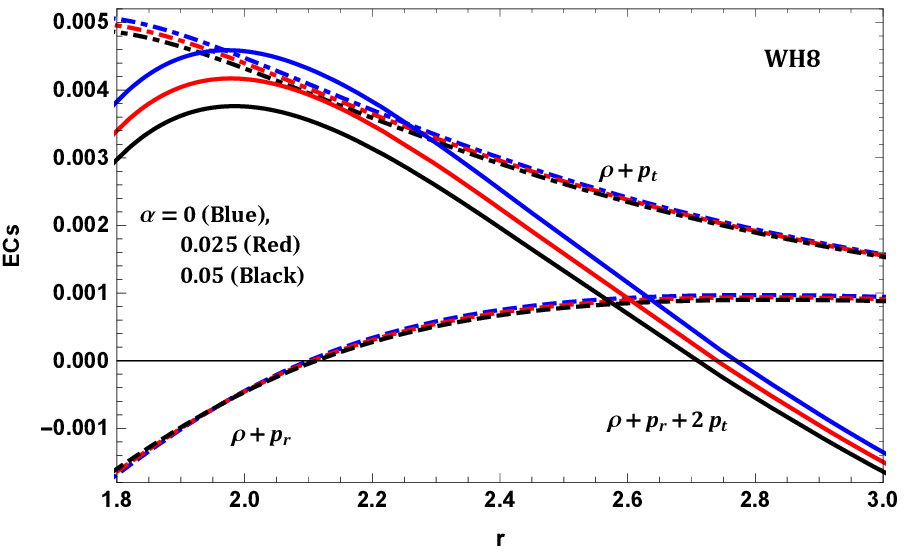}
\caption{Variation of energy conditions for WH8.}\label{f22}
\end{figure}

Using Eqs. (\ref{eq79}) and (\ref{eq34})-(\ref{eq36}), we can obtain the energy density and pressure components as
\begin{eqnarray}
&&\rho = \frac{1}{16 \pi  r^8} \Big[B r^8-12 a (r-r_0) \Big\{3 r^3-3 r^2 (2 \alpha  r_0+r_0)  \nonumber \\
&& \hspace{5 mm} +\alpha (3 \alpha +4) r r_0^2-\alpha ^2 r_0^3\Big\} \Big], \label{eq80}\\
&&\rho + p_r = \frac{3 a (r-r_0)}{2 \pi  r^8} \Big[2 r^3-(5 \alpha +3) r^2 r_0 \nonumber \\
&& \hspace{5 mm} +\alpha  (3 \alpha +7) r r_0^2-4 \alpha ^2 r_0^3 \Big], \\
&&\rho + p_t = \frac{3 a (r-r_0) (r-\alpha  r_0)}{2 \pi  r^6}, \\
&&\rho + p_r + 2p_t =  \frac{1}{8 \pi  r^8} \Big[12 a (r-r_0) \Big\{7 r^3-(13 \alpha +6) r^2 r_0+ \nonumber \\
&& \hspace{5 mm} \alpha  (6 \alpha +11) r r_0^2-5 \alpha ^2 R^3\Big\}-B r^8 \Big].\label{eq83}
\end{eqnarray}

\begin{figure}[t]
\centering
\includegraphics[scale=.8]{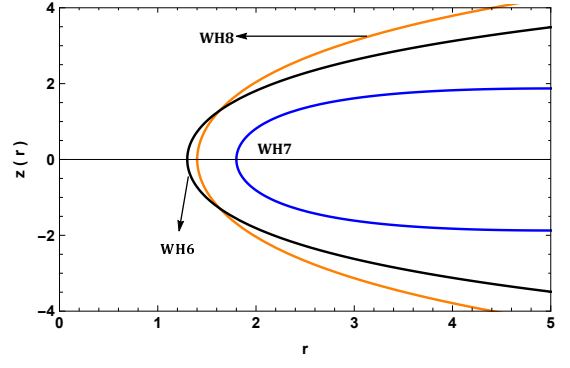}
\caption{Plots of the embedded surface $z(r)$ for WH6, WH7 and WH8.}\label{f23}
\end{figure}

\begin{figure}[t]
\centering
\includegraphics[scale=.75]{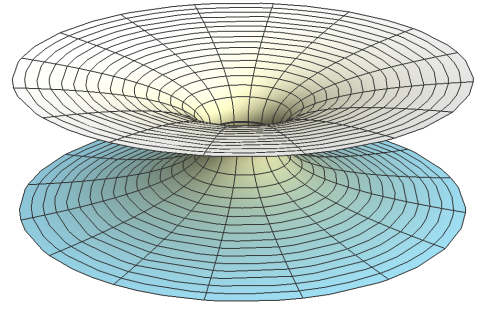}
\caption{The embedding diagram for WH5.}\label{f24}
\end{figure}

\begin{figure}[t]
\centering
\includegraphics[scale=.75]{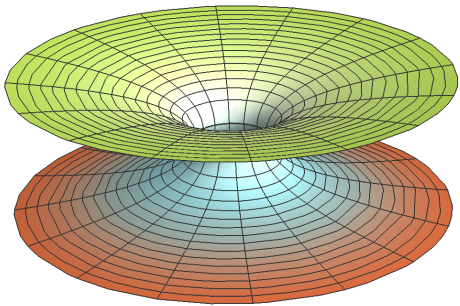}
\caption{The wormhole embedding for WH8.}\label{f25}
\end{figure}
Based on Eq. (\ref{eq79}) and (\ref{c4}), the behaviors of $b(r)$, $b(r)-r$, $b'(r)$ and $b(r)/r$ are displayed in Fig. \ref{f19}. The $b(r)-r$ cuts the $r$-axis at $r_0=1.4$, which is the throat of the wormhole WH8.

According to Eqs. (\ref{eq80})-(\ref{eq83}) we discuss about the energy conditions. The corresponding results for the pressure and density profile are shown in Figs. \ref{f21} and \ref{f22}, respectively. As mentioned in the above discussion, one can find that $\rho >0$ inside and outside the throat for the specific value of $a=0.5, r_0=1.4, c_3=1.165$,  $B=0.2$ and for different values of $\alpha$. In this analogy, the energy density has regions of positive magnitude near the throat, and regions with negative radial pressure which tend to zero from above in the asymptotic region.  Our results show that NEC is violated when $r \geq r_0$ and hence the WEC is violated also. The other embedding surface $z(r)$ requires numerical integration, and for that purpose we use ``NIntegrate" package which is depicted in Fig. \ref{f23}. Fig. \ref{f25} shows the revolution surface.

\section{Results and discussions}\label{sec8}

In this paper, we have discussed wormhole geometries in the context of teleparallel equivalent of general relativity (TEGR) and its straightforward extension of $f(T)$ gravity.  Since the teleparallel models of gravity are based on the torsion tensor while GR is formulated using the curvature. 
Motivated by the attempts to explain the observed late
time accelerated expansion of the universe, $f(T)$ theories of gravity have been extensively applied to cosmology without invoking the dark energy.  An important argument is that when $f(T)=T$ we recover the well-known conservation equation of TEGR. However, in the pure tetrad formalism,
$f(T)$ gravity exhibits violation of the local Lorentz invariance.

In this paper, we have discussed wormhole configuration in TEGR and $f(T)$ context. In particular, wormhole physics possess a peculiar property, namely \textit{exotic matter}, involving a stress-energy tensor that violates the null energy condition. We have developed the wormhole solution under the assumption of spherical symmetry and the existence of a conformal symmetries.  Since, we know that if a spacetime admits conformal symmetry then there exists a conformal killing vector field in the spacetime, which reduces the number unknown quantities also.
We have studied various type of solutions with the exotic matter restricted at the throat neighborhood applying the cut-and-paste approach of the stress-energy tensor at a junction interface. This approach is motivated for finding asymptotically flat geometries.

For this purpose, we explore wormhole solutions in TEGR gravity, and shows that stress-energy tensor violates the null energy condition (see Eq. (\ref{eq33})) to maintain the flaring out condition. Furthermore, we also consider phantom energy EoS, which violates the null energy condition. In this manner, in TEGR, exact solution was found for the  case of $\omega \neq -1, -3$, and the interval $-3 < \omega < -1$.  An interesting feature of the  phantom regime is that $\rho > 0$ throughout the spacetime. More specifically, for the case of 
phantom wormholes, it was found that infinitesimal amounts of phantom energy may support traversable wormholes. By carefully constructing specific  and different shape functions, we have analysed the wormhole geometries and discussed some of the properties of the resulting spacetime.

In the second part of this article, is based on the power-law of $f(T)$ model. Considering the field equations with a diagonal tetrad and the anisotropic fluid matter distribution,
a plethora of asymptotically flat exact solutions were found for different shape functions. This analysis shows that NEC is identically zero at the throat $r_0$ i.e. $(\rho+p_r)|_{r_0}=0$ (see Eq. \ref{eq38}). However, the energy density is positive inside and outside the throat radius. One important property of the solutions is that the matter obeys the NEC at the throat, but outside the throat radius NEC is violated and goes upto to the radius $r \leq R$. This situation is quite different from TEGR solution, but not new in wormhole physics. Since, the redshift function is not finite when $r \rightarrow \infty$ due to the conformal symmetry. Therefore, one needs a cut-off of the stress-energy by matching the interior solution to an exterior vacuum spacetime, at a junction interface.

Thus, it is safe to conclude that for the choice of diagonal tetrad, we found several solutions of wormhole geometries that violate the NEC at the throat and its neighbourhood in both TEGR and $f(T)$ gravity theories. However, the wormhole geometries in teleparallel gravity is more appealing than the $f(T)$ gravity.

\subsection*{Acknowledgments}
Farook Rahaman  would like to thank the authorities of the Inter-University Centre for Astronomy and Astrophysics, Pune, India for providing the research facilities.  FR is  also thankful to DST-SERB,  Govt. of India and RUSA 2.0, Jadavpur University,  for financial support.


\begin{thebibliography}{90}
\bibitem{Einstein} A. Einstein and N. Rosen, Phys. Rev.  \textbf{48}, 73 (1935).

\bibitem{Wheeler} J. A. Wheeler,  Geons Phys. Rev. {\bf 97} 511 (1955).

\bibitem{Wheeler1}   D. R. Brill and J. A. Wheeler,  Rev. Mod. Phys. {\bf 29}, 465 (1957).

\bibitem{Hartle}   D. R. Brill  and J. B.  Hartle, Phys. Rev. {\bf 135}, B271–8  (1964).

\bibitem{Wheeler2} J. A. Wheeler, Geometrodynamics, (Academic Press, New York, 1962).

\bibitem{whel} J. Wheeler, Ann. Phys. \textbf{2}, (6) 604 (1957).

\bibitem{Fuller1962} 
R. W. Fuller and J. A. Wheeler
Phys. Rev. \  {\bf 128}, 919 (1962).



\bibitem{Morris:1988cz} 
  M.~S.~Morris and K.~S.~Thorne,
  Am.\ J.\ Phys.\  {\bf 56}, 395 (1988).

\bibitem{Morris:1988tu} 
  M.~S.~Morris, K.~S.~Thorne and U.~Yurtsever,
  Phys.\ Rev.\ Lett.\  {\bf 61}, 1446 (1988).

\bibitem{Visser}  M. Visser, \textit{Lorentzian Wormholes: From Einstein to Hawking} (American Institute of Physics, New York, 1995).

\bibitem{Ori1993}   
A. Ori
Phys. Rev. Lett. {\bf 71}, 2517 (1993).

\bibitem{Ori1994}
  A. Ori and Y. Soen
Phys. Rev. D {\bf 49}, 3990 (1994).

\bibitem{Wang:1995nj}
  A.~Wang and P.~S.~Letelier,
  Prog.\ Theor.\ Phys.\  {\bf 94}, 137 (1995).

\bibitem{Visser:2003yf}
  M.~Visser, S.~Kar and N.~Dadhich,
   Phys.\ Rev.\ Lett.\   {\bf 90}, 201102 (2003).
  
  \bibitem{Kuhfittig:2002ur}
  P.~K.~F.~Kuhfittig,
   Phys.\ Rev.\ D {\bf 66},  024015 (2002).

  \bibitem{Kuhfittig:1999ur}
P. K. F. Kuhfittig, 
 Am.\ J.\ Phys. {\bf 67}, 125 (1999).  

\bibitem{Kar:1995ss}
  S.~Kar and D.~Sahdev,
  Phys.\ Rev.\ D {\bf 53}, 722 (1996).
  
  \bibitem{Kar:1994tz}
  S.~Kar,
  Phys.\ Rev.\ D {\bf 49}, 862 (1994).

 \bibitem{Visser:1989kh}
  M.~Visser,
   Phys.\ Rev.\ D {\bf 39}, 3182 (1989).

  \bibitem{Visser:1989kg}
  M.~Visser,
   Nucl.\ Phys.\ B {\bf 328}, 203 (1989).
  
  \bibitem{Lobo:2009ip}
  F.~S.~N.~Lobo and M.~A.~Oliveira,
  Phys.\ Rev.\ D {\bf 80}, 104012 (2009).
  
  \bibitem{Mazharimousavi:2012xv}
  S.~Habib Mazharimousavi and M.~Halilsoy,
  Mod.\ Phys.\ Lett.\ A {\bf 31}, no. 37, 1650203 (2016).
  
  \bibitem{Pavlovic:2014gba}
  P.~Pavlovic and M.~Sossich,
  Eur.\ Phys.\ J.\ C {\bf 75}, 117 (2015).
  
  \bibitem{Sharif:2018jdj}
  M.~Sharif and I.~Nawazish,
  Annals Phys.\  {\bf 389}, 283 (2018).
  
  \bibitem{DeBenedictis:2012qz}
  A.~DeBenedictis and D.~Horvat,
  Gen.\ Rel.\ Grav.\  {\bf 44}, 2711 (2012)

\bibitem{MontelongoGarcia:2010xd}
  N.~Montelongo Garcia and F.~S.~N.~Lobo,
  Class.\ Quant.\ Grav.\  {\bf 28}, 085018 (2011).
  
  \bibitem{Bhawal:1992sz}
  B.~Bhawal and S.~Kar,
  Phys.\ Rev.\ D {\bf 46}, 2464 (1992).
  
  \bibitem{Mehdizadeh:2015jra}
  M.~R.~Mehdizadeh, M.~Kord Zangeneh and F.~S.~N.~Lobo,
  Phys.\ Rev.\ D {\bf 91},  084004 (2015).
  
  \bibitem{Kanti:2011yv}
  P.~Kanti, B.~Kleihaus and J.~Kunz,
  Phys.\ Rev.\ D {\bf 85}, 044007 (2012).
  
  \bibitem{Maeda:2008nz}
  H.~Maeda and M.~Nozawa,
  Phys.\ Rev.\ D {\bf 78}, 024005 (2008).
  
  \bibitem{Shaikh:2018yku}
  R.~Shaikh,
  Phys.\ Rev.\ D {\bf 98},  064033 (2018).
  
  \bibitem{Dehghani:2009zza}
  M.~H.~Dehghani and Z.~Dayyani,
  Phys.\ Rev.\ D {\bf 79}, 064010 (2009).
  
  \bibitem{Zangeneh:2015jda}
  M.~Kord Zangeneh, F.~S.~N.~Lobo and M.~H.~Dehghani,
  Phys.\ Rev.\ D {\bf 92}, 124049 (2015).
  
  \bibitem{Matulich:2011ct}
  J.~Matulich and R.~Troncoso,
  JHEP {\bf 1110}, 118 (2011).
  
   \bibitem{Moraes:2017mir}
  P.~H.~R.~S.~Moraes and P.~K.~Sahoo,
   Phys.\ Rev.\ D {\bf 96}, 044038 (2017).
  
  \bibitem{Elizalde:2018frj}
  E.~Elizalde and M.~Khurshudyan,
  Phys.\ Rev.\ D {\bf 98},  123525 (2018).
  
  \bibitem{Elizalde:2018arz}
  E.~Elizalde and M.~Khurshudyan,
  Phys.\ Rev.\ D {\bf 99},  024051 (2019).
  
  \bibitem{Banerjee:2019wjj}
  A.~Banerjee, K.~N.~Singh, M.~K.~Jasim and F.~Rahaman,
  \url{arXiv:1908.04754 [gr-qc]}.
  
  \bibitem{Unzicker:2005tz}
A. Unzicker and T. Case
    \url{arXiv:physics/0503046 [physics.hist-ph]}.
    
    \bibitem{Shirafuji:1995xc}
  T.~Shirafuji, G.~G.~L.~Nashed and K.~Hayashi,
  Prog.\ Theor.\ Phys.\  {\bf 95}, 665 (1996).
  
  \bibitem{Maluf:2013gaa}
  J.~W.~Maluf,
  Annalen Phys.\  {\bf 525}, 339 (2013).

\bibitem{Okolow:2013ifa}
  A.~Okolow,
  Gen.\ Rel.\ Grav.\  {\bf 46}, 1653 (2014).
  
  \bibitem{Ortin}
   T. Ortin, 
   Cambridge:  Cambridge University Press (2015).
  
  \bibitem{Ferraro:2006jd}
  R.~Ferraro and F.~Fiorini,
  Phys.\ Rev.\ D {\bf 75}, 084031 (2007).
  
  \bibitem{Ferraro:2008ey}
  R.~Ferraro and F.~Fiorini,
  Phys.\ Rev.\ D {\bf 78}, 124019 (2008).

\bibitem{Cai:2015emx}
  Y.~F.~Cai, S.~Capozziello, M.~De Laurentis and E.~N.~Saridakis,
  Rept.\ Prog.\ Phys.\  {\bf 79},  106901 (2016).
  
 \bibitem{Krssak:2015oua}
  M.~Krššák and E.~N.~Saridakis,
  Class.\ Quant.\ Grav.\  {\bf 33},  115009 (2016).

\bibitem{Golovnev:2017dox}
  A.~Golovnev, T.~Koivisto and M.~Sandstad,
  Class.\ Quant.\ Grav.\  {\bf 34},  145013 (2017).
  
 \bibitem{Bejarano:2019fii} 
  C.~Bejarano, R.~Ferraro, F.~Fiorini and M.~J.~Guzmán,
  Universe {\bf 5}, 158 (2019).
  
  
  \bibitem{Tamanini:2012hg}
  N.~Tamanini and C.~G.~Boehmer,
  Phys.\ Rev.\ D {\bf 86}, 044009 (2012).
  
  \bibitem{Wu:2010xk}
  P.~Wu and H.~W.~Yu,
  Phys.\ Lett.\ B {\bf 692}, 176 (2010).
  
  \bibitem{Wu:2010mn}
  P.~Wu and H.~W.~Yu,
  Phys.\ Lett.\ B {\bf 693}, 415 (2010).
  
  \bibitem{Myrzakulov:2010vz}
  R.~Myrzakulov,
  Eur.\ Phys.\ J.\ C {\bf 71}, 1752 (2011).
  
  \bibitem{Karami:2013rda}
  K.~Karami and A.~Abdolmaleki,
  Res.\ Astron.\ Astrophys.\  {\bf 13}, 757 (2013).
  
  \bibitem{Chen:2010va}
  S.~H.~Chen, J.~B.~Dent, S.~Dutta and E.~N.~Saridakis,
  Phys.\ Rev.\ D {\bf 83}, 023508 (2011).
  
  \bibitem{Dent:2011zz}
  J.~B.~Dent, S.~Dutta and E.~N.~Saridakis,
  JCAP {\bf 1101}, 009 (2011).
  
  \bibitem{Wang:2011xf}
  T.~Wang,
  Phys.\ Rev.\ D {\bf 84}, 024042 (2011).
  
  \bibitem{Iorio:2012cm}
  L.~Iorio and E.~N.~Saridakis,
  Mon.\ Not.\ Roy.\ Astron.\ Soc.\  {\bf 427}, 1555 (2012).
  
  \bibitem{Deliduman:2011ga}
  C.~Deliduman and B.~Yapiskan,
  \url{arXiv:1103.2225 [gr-qc]}.
  
  \bibitem{Wu:2011xa}
  P.~Wu and H.~Yu,
  Phys.\ Lett.\ B {\bf 703}, 223 (2011).
  
  \bibitem{Nashed:uja}
  G.~G.~L.~Nashed,
  Gen.\ Rel.\ Grav.\  {\bf 45}, 1887 (2013).
  
 \bibitem{Boehmer:2011gw} 
  C.~G.~Boehmer, A.~Mussa and N.~Tamanini,
  Class.\ Quant.\ Grav.\  {\bf 28}, 245020 (2011).
  
\bibitem{Singh:2019ykp}
  K.~N.~Singh, F.~Rahaman and A.~Banerjee,
  Phys.\ Rev.\ D {\bf 100},  084023 (2019).

  \bibitem{Das:2015gwa}
  A.~Das, F.~Rahaman, B.~K.~Guha and S.~Ray,
  Astrophys.\ Space Sci.\  {\bf 358}, 36 (2015).

  \bibitem{Chanda:2019hyh}
  A.~Chanda, S.~Dey and B.~C.~Paul,
  Eur.\ Phys.\ J.\ C {\bf 79},  502 (2019).

\bibitem{Abbas:2015xia}
  G.~Abbas, S.~Qaisar and M.~A.~Meraj,
  Astrophys.\ Space Sci.\  {\bf 357},  156 (2015).
  
  \bibitem{Bohmer:2011si}
  C.~G.~Boehmer, T.~Harko and F.~S.~N.~Lobo,
  Phys.\ Rev.\ D {\bf 85}, 044033 (2012)
  
  \bibitem{Jamil:2012ti}
  M.~Jamil, D.~Momeni and R.~Myrzakulov,
  Eur.\ Phys.\ J.\ C {\bf 73}, 2267 (2013)

\bibitem{Lin:2019tyw}
  R.~H.~Lin, Z.~Y.~Wu and X.~H.~Zhai,
  \url{arXiv:1906.10323 [gr-qc]}.
  
  \bibitem{Sharif:2013exa}
  M.~Sharif and S.~Rani,
  Phys.\ Rev.\ D {\bf 88},  123501 (2013).
  
  \bibitem{Sharif:2013lya}
  M.~Sharif and S.~Rani,
  Gen.\ Rel.\ Grav.\  {\bf 45}, 2389 (2013).
  
  \bibitem{Boehmer:2007md}
  C.~G.~Boehmer, T.~Harko and F.~S.~N.~Lobo,
  Class.\ Quant.\ Grav.\  {\bf 25}, 075016 (2008).
  
\bibitem{Herrera1984}  L. Herrera, J. Jimenez, L. Leal, J. Ponce de Leon and M. Esculpi and V. Galina,,  
J.\ Math.\ Phys.\  {\bf 25}, 3274 (1984). 

\bibitem{Herrera1985}
L. Herrera and J. Ponce de Leon, 
J.\ Math.\ Phys.\  {\bf 26}, 2303 (1985).

\bibitem{Maartens} R. Maartens and M. S. Maharaj, J.\ Math.\ Phys.\  {\bf 31}, 151 (1990).

\bibitem{Mak2004}
M. K. Mak and T. Harko
Int.\ J.\ Mod.\ Phys.\ D {\bf  13}, 149 (2004).






\bibitem{Kuhfittig:2015cea}
  P.~K.~F.~Kuhfittig,
  Annals Phys.\  {\bf 355}, 115 (2015).
  
  \bibitem{Rahaman:2013ywa}
  F.~Rahaman, S.~Ray, G.~S.~Khadekar, P.~K.~F.~Kuhfittig and I.~Karar,
  Int.\ J.\ Theor.\ Phys.\  {\bf 54}, 699 (2015).
  
  \bibitem{Bhar:2016vdn}
  P.~Bhar, F.~Rahaman, T.~Manna and A.~Banerjee,
  Eur.\ Phys.\ J.\ C {\bf 76}, 708 (2016).
  
  \bibitem{Sharif:2016aom}
  M.~Sharif and H.~I.~Fatima,
  Gen.\ Rel.\ Grav.\  {\bf 48}, 148 (2016).
  
  \bibitem{Momeni:2016oai} 
  D.~Momeni, G.~Abbas, S.~Qaisar, Z.~Zaz and R.~Myrzakulov,
  Can.\ J.\ Phys.\  {\bf 96}, 1295 (2018).
  
  \bibitem{Abbas:2015zua} 
  G.~Abbas, S.~Qaisar, A.~Jawad, S.~Qaisar and A.~Jawad,
  Astrophys.\ Space Sci.\  {\bf 359}, 57 (2015).
  

\bibitem{Arellano:2006np}
  A.~V.~B.~Arellano and F.~S.~N.~Lobo,
  Class.\ Quant.\ Grav.\  {\bf 23}, 7229 (2006).
  
  \bibitem{Poisson:1995sv}
  E.~Poisson and M.~Visser,
  Phys.\ Rev.\ D {\bf 52}, 7318 (1995).

\bibitem{Velay-Vitow:2017odc}
  J.~Velay-Vitow and A.~DeBenedictis,
  Phys.\ Rev.\ D {\bf 96}, 024055 (2017).


\bibitem{delaCruz-Dombriz:2014zaa}
  A.~de la Cruz-Dombriz, P.~K.~S.~Dunsby and D.~Saez-Gomez,
  JCAP {\bf 1412}, 048 (2014).
  
  \bibitem{Rahaman:2014dpa}
  F.~Rahaman, I.~Karar, S.~Karmakar and S.~Ray,
  Phys.\ Lett.\ B {\bf 746}, 73 (2015).
  
 
  
  \bibitem{Kuhfittig:2015xwa}
  P.~K.~F.~Kuhfittig,
  Eur.\ Phys.\ J.\ C {\bf 75},  357 (2015).
  
  \bibitem{Rahaman:2006xa}
  F.~Rahaman, M.~Kalam and S.~Chakraborty,
  Chin.\ J.\ Phys.\  {\bf 45},  518 (2007).

\end{thebibliography}
\end{document}